\documentclass[emulateapj]{emulateapj}
\usepackage{amsmath}
\usepackage{lineno}

\newcommand{\ktwo}{\emph{K2}}

\newcommand{\teff}{\mbox{$T_{\rm eff}$}}
\newcommand{\logg}{\mbox{$\log g$}}
\newcommand{\feh}{\mbox{$\rm{[Fe/H]}$}}
\newcommand{\fehval}{\mbox{$+0.42$}}

\newcommand{\rsun}{\mbox{$R_{\sun}$}}
\newcommand{\msun}{\mbox{$M_{\sun}$}}
\newcommand{\numax}{\mbox{$\nu_{\rm max}$}}
\newcommand{\dnu}{\mbox{$\Delta \nu$}}
\newcommand{\muHz}{\mbox{$\mu$Hz}}
\newcommand{\thestar}{\mbox{K2-97}}
\newcommand{\planrad}{\mbox{1.31 $\pm$ 0.11 R$_{\mathrm{J}}$}}
\newcommand{\planmass}{\mbox{0.48 $\pm$ 0.07 M$_{\mathrm{J}}$}}
\newcommand{\starrad}{\mbox{4.20 $\pm$ 0.14 R$_{\odot}$}}
\newcommand{\starmass}{\mbox{1.16 $\pm$ 0.12 M$_{\odot}$}}
\newcommand{\theotherstar}{\mbox{EPIC 228754001}}
\newcommand{\otherplanrad}{\mbox{1.30 $\pm$ 0.07 R$_{\mathrm{J}}$}}
\newcommand{\otherplanmass}{\mbox{0.49 $\pm$ 0.06 M$_{\mathrm{J}}$}}
\newcommand{\otherstarrad}{\mbox{3.85 $\pm$ 0.13 R$_{\odot}$}}
\newcommand{\otherstarmass}{\mbox{1.08 $\pm$ 0.08 M$_{\odot}$}}

\begin{document}



\bibliographystyle{apj} 


\title{Seeing double with K2: Testing Re-inflation with Two Remarkably Similar Planets Around Red Giant Branch Stars}

\author{Samuel K.\ Grunblatt\altaffilmark{1,*}}

\shorttitle{Planet Re-Inflation}
\shortauthors{Grunblatt et al.}

\author{Daniel Huber\altaffilmark{1,2,3,4}}
\author{Eric Gaidos\altaffilmark{5}}
\author{Eric D.\ Lopez\altaffilmark{6}}
\author{Andrew W.\ Howard\altaffilmark{1,7}}
\author{Howard T. Isaacson\altaffilmark{8}}
\author{Evan Sinukoff\altaffilmark{1,7}}
\author{Andrew Vanderburg\altaffilmark{9,10,16}}
\author{Larissa Nofi\altaffilmark{1,11}}
\author{Jie Yu\altaffilmark{2}}
\author{Thomas S. H. North\altaffilmark{12, 4}}
\author{William Chaplin\altaffilmark{12, 4}}
\author{Daniel Foreman-Mackey\altaffilmark{13,17}}
\author{Erik Petigura\altaffilmark{7,18}}
\author{Megan Ansdell\altaffilmark{1}}
\author{Lauren Weiss\altaffilmark{14,19}}
\author{Benjamin Fulton\altaffilmark{1,7,20}}
\author{Douglas N. C. Lin\altaffilmark{15}}

\altaffiltext{1}{Institute for Astronomy, University of Hawaii,
2680 Woodlawn Drive, Honolulu, HI 96822, USA}
\altaffiltext{2}{Sydney Institute for Astronomy (SIfA), School of Physics, University of 
Sydney, NSW 2006, Australia}
\altaffiltext{3}{SETI Institute, 189 Bernardo Avenue, Mountain View, CA 94043, USA}
\altaffiltext{4}{Stellar Astrophysics Centre, Department of Physics and Astronomy, 
Aarhus University, Ny Munkegade 120, DK-8000 Aarhus C, Denmark}
\altaffiltext{5}{Department of Geology $\&$ Geophysics, University of
Hawaii at Manoa, Honolulu, Hawaii 96822, USA} 
\altaffiltext{6}{NASA Goddard Space Flight Center, Greenbelt, MD 20771, USA}
\altaffiltext{7}{California Institute of Technology, Pasadena, CA 91125, USA}
\altaffiltext{8}{Department of Astronomy, UC Berkeley, Berkeley, CA 94720, USA}
\altaffiltext{9}{Harvard-Smithsonian Center for Astrophysics, 60 Garden St., Cambridge, MA 02138, USA}
\altaffiltext{10}{Department of Astronomy, The University of Texas at Austin, Austin, TX 78712, USA}
\altaffiltext{11}{Lowell Observatory, 1400 W. Mars Hill Road, Flagstaff, AZ 86001, USA}
\altaffiltext{12}{School of Physics and Astronomy, University of Birmingham, Birmingham, B15 2TT, United Kingdom}
\altaffiltext{13}{Astronomy Department, University of Washington, Seattle, WA}
\altaffiltext{14}{Institut de Recherche sur les Exoplan\`{e}tes, Universit\'{e} de
Montr\'{e}al, Montr\'{e}al, QC, Canada}
\altaffiltext{15}{UCO/Lick Observatory, Board of Studies in Astronomy and Astrophysics, University of California, Santa Cruz, California 95064, USA}
\altaffiltext{16}{NSF Graduate Research Fellow}
\altaffiltext{17}{Sagan Fellow}
\altaffiltext{18}{Hubble Fellow}
\altaffiltext{19}{Trottier Fellow}
\altaffiltext{20}{Texaco Fellow}
\altaffiltext{*}{skg@ifa.hawaii.edu}

\begin{abstract}


Despite more than 20 years since the discovery of the first gas giant planet with an anomalously large radius, the mechanism for planet inflation remains unknown. Here, we report the discovery of EPIC~228754001.01, an inflated gas giant planet found with the NASA \emph{K2} Mission, and a revised mass for another inflated planet, K2-97b. These planets reside on $\approx$9~day orbits around host stars which recently evolved into red giants. We constrain the irradiation history of these planets using models constrained by asteroseismology and Keck/HIRES spectroscopy and radial velocity measurements. We measure planet radii of \planrad\ and \otherplanrad{}, respectively. These radii are typical for planets receiving the current irradiation, but not the former, zero age main sequence irradiation of these planets. This suggests that the current sizes of these planets are directly correlated to their current irradiation. Our precise constraints of the masses and radii of the stars and planets in these systems allow us to constrain the planetary heating efficiency of both systems as 0.03\%$^{+0.03\%}_{ -0.02\%}$. These results are consistent with a planet re-inflation scenario, but suggest the efficiency of planet re-inflation may be lower than previously theorized. Finally, we discuss the agreement within 10\% of stellar masses and radii, and planet masses, radii, and orbital periods of both systems and speculate that this may be due to selection bias in searching for planets around evolved stars.  

\end{abstract}

\section{Introduction}

Since the first measurement of planet radii outside our Solar System \citep{charbonneau2000, henry2000}, it has been known that gas giant planets with equilibrium temperatures greater than 1000 K tend to have radii larger than model predictions \citep{burrows1997, bodenheimer2001, guillot2002}. Moreover, a correlation has been observed between incident stellar radiation and planetary radius inflation \citep{burrows2000, laughlin2011, lopez2016}. The diversity of mechanisms proposed to explain the inflation of giant planets \citep{baraffe2014} can be split into two general classes: mechanisms where stellar irradiation is deposited directly into the planet's deep interior, driving adiabatic heating of the planet and thus inflating its radius \citep[Class I, e.g.,][]{bodenheimer2001, batygin2010, ginzburg2016}, and mechanisms where no energy is deposited into the deep planetary interior and the inflationary mechanism simply acts to slow radiative cooling of the planet's atmosphere, preventing it from losing its initial heat and thus radius inflation from its formation \citep[Class II, e.g.,][]{burrows2000, chabrier2007, wu2013}. These mechanism classes can be distinguished by measuring the radii of planets that have recently experienced a large changes in irradiation, such as planets orbiting red giant stars at 10-30 day orbital periods \citep{lopez2016}. To quantify the distinction between mechanism classes, we require that planets 1) approach or cross the empirical planet inflation threshold of 2$\times$10$^{8}$~erg s$^{-1}$ cm$^{-2}$ \citep[$\approx$150 F$_\oplus$][]{demory2011}) after reaching the zero-age main sequence, and 2) experience a change in incident flux large enough that the planet radius would increase significantly, assuming it followed the trend between incident flux and planet radius found by \citet{laughlin2011}. If such planets are currently inflated, heat from irradiation must have been deposited directly into the planet interior, indicating that Class I mechanisms must be at play, whereas if these planets are not inflated, no energy has been transferred from the planet surface into its deep interior, and thus Class II mechanisms are favored. By constraining the efficiency of heat transfer to inflated planets orbiting evolved host stars, we can distinguish the efficiency of these two classes of inflation mechanisms \citep{lopez2016, ginzburg2016}.



To constrain the properties of giant planet inflation, we search for transiting giant planets orbiting low luminosity red giant branch (LLRGB) stars with the NASA \emph{K2} Mission \citep{howell2014, huber15}. These stars are large enough that we can detect their oscillations to perform asteroseismology but small enough that gas giant planet transits are still detectable in \emph{K2} long cadence data. Close-in planets in these systems have experienced significant changes in irradiation over time. The first planet discovered by our survey, \thestar{b}, was published by \citet[][hereafter referred to as G16]{grunblatt2016}. Using a combination of asteroseismology, transit analysis and radial velocity measurements, G16 measured the mass and radius of this planet to be 1.10 $\pm$ 0.12 M$_\mathrm{J}$ and \planrad{}, respectively. This implied a direct heating efficiency of 0.1\%--0.5\%, suggesting that the planet radius was directly influenced by the increase in irradiation caused by the host star evolution. 

Here, we present additional radial velocity data that revise the mass of \thestar{} to \planmass, as well as the discovery of the second planet in our survey, \theotherstar{.01}, with a radius of \otherplanrad\ and mass of \otherplanmass. These planets currently receive incident fluxes between 700 and 1100 F$_\oplus$, but previously received fluxes between 100 and 350 F$_\oplus$ when the host stars were on the main sequence. Quantifying the incident flux evolution of these systems allows us to estimate the planetary heating efficiency and distinguish between planetary inflation mechanisms. 


\section{Observations}

\subsection{K2 Photometry}
 
\begin{figure*}[ht!]
\epsscale{1.0}
\plotone{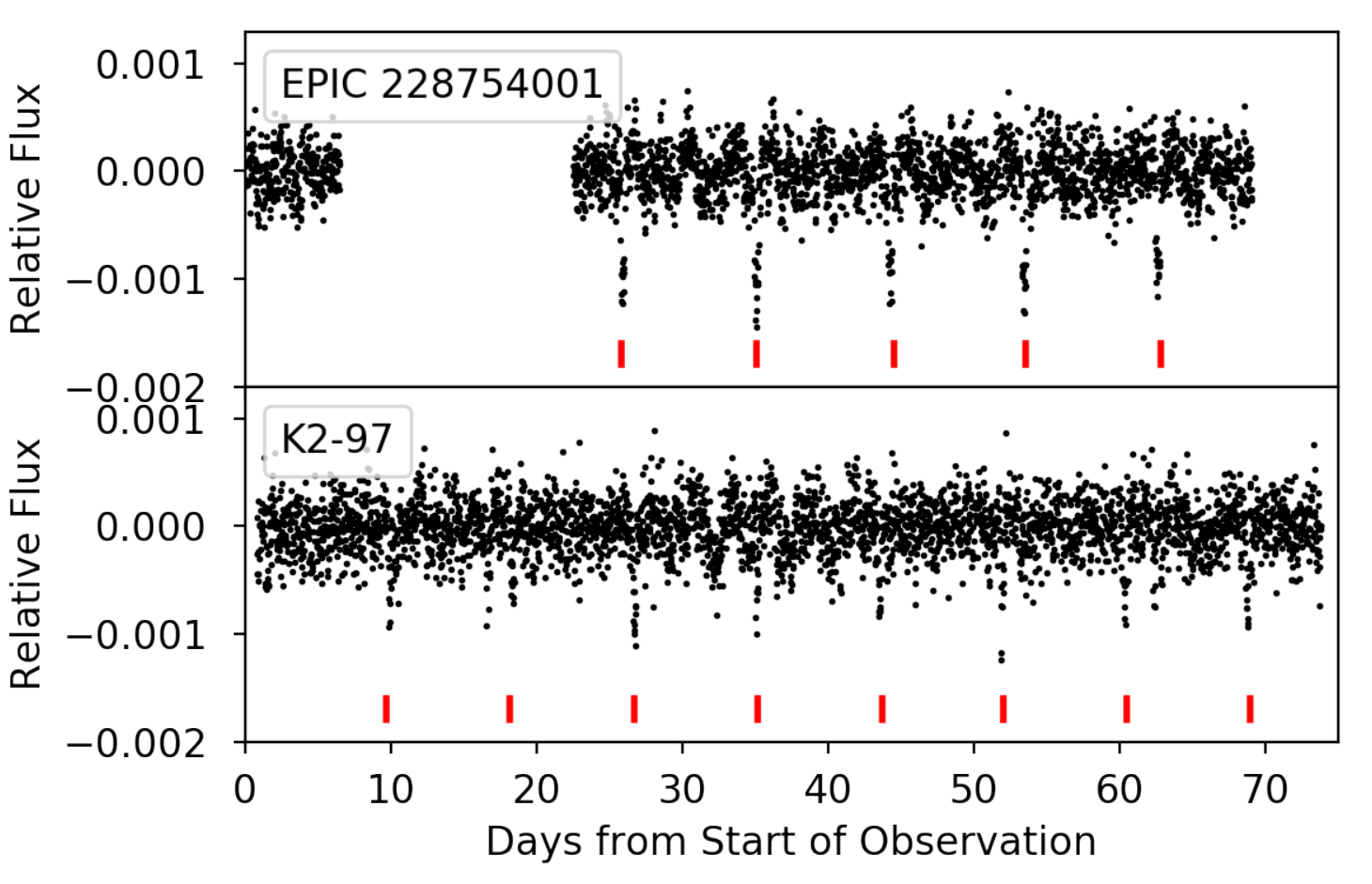}
\caption{Detrended K2 lightcurves of \thestar{} (bottom) and \theotherstar{} (top). These lightcurves were produced using a modified method of the pipeline presented in \citet{vanderburg2016}, where both instrument systematics and the planet transit were modeled simultaneously to prevent transit dilution. The lightcurve has been normalized and median filtered as well as unity subtracted. Individual transits are visible by eye, and are denoted by red fiducial marks. \label{lightcurve}}
\end{figure*}

 In the \emph{K2} extension to the NASA \emph{Kepler} mission, multiple fields along the ecliptic are observed almost continuously for approximately 80 days \citep{howell2014}. EPIC 211351816 (now known as K2-97; G16) was selected for observation as a part of \emph{K2} Guest Observer Proposal GO5089 (PI: Huber) and observed in Campaign 5 of \emph{K2} during the first half of 2015. EPIC 228754001 was selected and observed in Campaign 10 of \emph{K2} as part of \emph{K2} Guest Observer Proposal GO10036 (PI: Huber) in the second half of 2016. As the \emph{Kepler} telescope now has unstable pointing due to the failure of two of its reaction wheels, it is necessary to correct for the pointing-dependent error in the flux received per pixel. We produced a lightcurve by simultaneously fitting thruster systematics, low frequency variability, and planet transits with a Levenberg-Marquardt minimization algorithm, using a modified version of the pipeline from \citet{vanderburg2016}. These lightcurves were then normalized and smoothed with a 75 hour median filter, and points deviating from the mean by more than 5-$\sigma$ were removed. By performing a box least-squares transit search for transits with 5- to 40-day orbital periods and 3- to 30-hr transit durations on these lightcurves using the algorithm of \citet{kovacs2002}, we identified transits of $\approx$500 and $\approx$1000 ppm, respectively. Using the techniques of G16 and those described in \S 4.1, we determined the transits came from an object which was planetary in nature. Figure \ref{lightcurve} shows our adopted lightcurves for \thestar{} and \theotherstar{}.

\subsection{Imaging with Keck/NIRC2 AO}

To check for potential blended background stars, we obtained natural guide-star adaptive optics (AO) images of \theotherstar{} through the broad $K'$ filter ($\lambda_{\mathrm{center}}$ = 2.124 $\mu$m) with the Near-Infrared Camera (NIRC2) at the Keck-2 telescope on Maunakea during the night of UT 25 January 2017. The narrow camera (pixel scale 0.01") was used for all sets of observations.  No additional sources were detected within $\sim$3"  of the star.  The contrast ratio of the detection limit is more than 7 magnitudes at 0.5"; brighter objects could be detected to within 0.15" of the star. These data were collected to quantify the possibility of potential false positive scenarios in these systems, and the relevant analysis is described in \S 4.2. Previous analysis by G16 of NIRC2 AO images of \thestar{} reached effectively identical conclusions.

\begin{figure}[ht!]
\epsscale{1.2}
\plotone{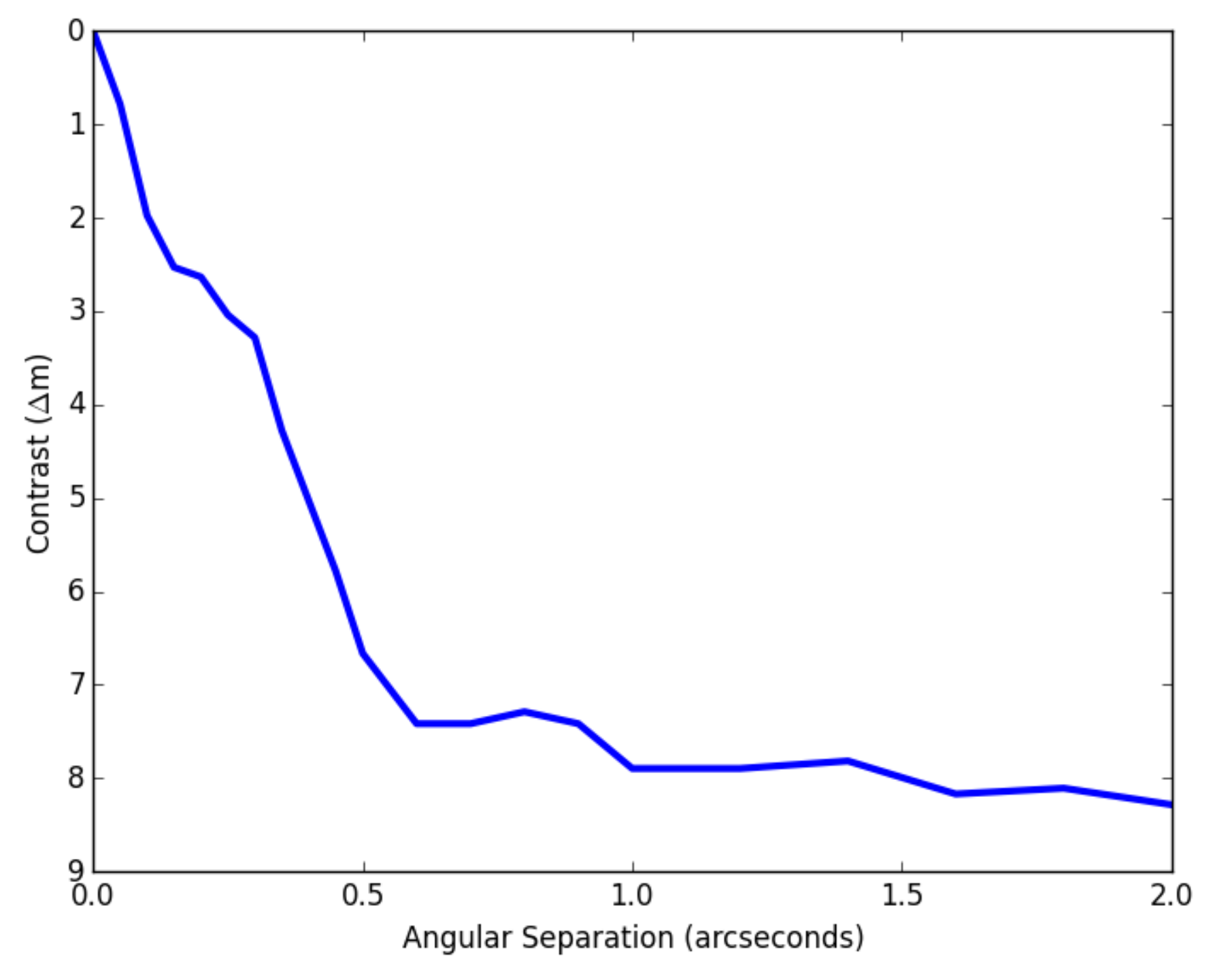}
\caption{Contrast in differential $K'$ magnitude as a function of angular separation from \theotherstar{}. No companions were detected within 3" of the source. G16 found effectively identical results for \thestar{}. \label{contrastcurve}}
\end{figure}

Images were processed using a custom Python pipeline that linearized, dark-subtracted, flattened, sky-subtracted, and co-added the images \citep{metchev2009}. A cutout $\sim$3.0'' across, centered on the star, was made and inserted back into the processed image as a simulated companion. A contrast curve was generated by decreasing the brightness and angular separation of the simulated companion with respect to the primary, until the limits of detection (3.0$\sigma$) were reached. Figure \ref{contrastcurve} plots the contrast ratio for detection as a function of distance from the source \theotherstar{}.

\subsection{High-Resolution Spectroscopy and Radial Velocity Measurements with Keck/HIRES}

We obtained a high resolution, high signal-to-noise spectrum of \thestar{} and \theotherstar{} using the High Resolution Echelle Spectrometer (HIRES) on the 10 meter Keck-I telescope at Mauna Kea Observatory on the Big Island of Hawaii. HIRES provides spectral resolution of roughly 65,000 in a wavelength range of 4500 to 6200 $\mathrm{\AA}$ \citep{vogt1994}, and has been used to both characterize over 1000 \emph{Kepler} planet host stars \citep{petigura2017} as well as confirm and provide precise parameters of over 2000 \emph{Kepler} planets \citep{fulton2017, johnson2017}. Our spectra were analyzed using the software package SpecMatch \citep{petigura2015} following the procedure outlined in G16.


Radial velocity (RV) measurements were obtained between January 27, 2016 and April 10, 2017 using the High Resolution Echelle Spectrometer (HIRES) on the Keck-I Telescope at the Mauna Kea Observatory in Hawaii. Individual measurements are listed in Table \ref{tbl-rv} and shown in Figure \ref{rv}. All RV spectra were obtained through an iodine gas cell. We collected three measurements of \thestar{} with Keck/HIRES in 2016, and seven additional measurements in 2017. All eleven measurements of \theotherstar{} were taken between December 2016 and April 2017. Fits to the radial velocity data were made using the publicly available software package \texttt{RadVel} \citep{radvel} and confirmed through independent analysis presented in \S 4.2. We adopted the same method for radial velocity analysis as described in G16 \citep{butler1996}.

\begin{deluxetable}{ccrrrr}
\tabletypesize{\scriptsize}
\tablecaption{Radial Velocities\label{tbl-rv}}
\tablewidth{0pt}
\tablehead{
\colhead{Star} & \colhead{BJD-2440000} & \colhead{RV (m s$^{-1}$)} & \colhead{Prec. (m s$^{-1}$)}
}
\startdata
K2-97 & 17414.927751 & -4.91 & 1.79 \\
K2-97 & 17422.855362 & -38.94 & 1.72 \\
K2-97 & 17439.964043 & -17.95 & 2.22 \\
K2-97 & 17774.905553 & -44.03 & 1.85 \\
K2-97 & 17790.840786 & -50.74 & 1.77 \\
K2-97 & 17802.819367 & 7.96  & 1.76 \\
K2-97 & 17803.836621 & 38.90 & 1.64 \\
K2-97 & 17830.802784 & 32.84 & 1.77\\
K2-97 & 17853.790069 & 23.05 & 1.78 \\
K2-97 & 17854.774479 & 46.68 & 1.85 \\

EPIC228754001 & 17748.099507 & -30.32 & 1.95 \\
EPIC228754001 & 17764.115738 & 25.80 & 1.66 \\
EPIC228754001 & 17766.139232 & -40.85 & 1.96 \\
EPIC228754001 & 17776.065142 & -26.91 & 1.54 \\
EPIC228754001 & 17789.093812 & 26.09 & 1.74 \\
EPIC228754001 & 17790.091515 & 45.40 & 1.68 \\
EPIC228754001 & 17791.071462 & 46.31 & 1.85 \\
EPIC228754001 & 17794.992775 & -22.43 & 1.88 \\
EPIC228754001 & 17803.927316 & -37.99 & 1.91 \\
EPIC228754001 & 17830.066681 & -34.92 & 1.83 \\
EPIC228754001 & 17854.937650 & 50.42 & 1.78 \\

\enddata 
\tablecomments{The precisions listed here are instrumental only, and do not take into account the uncertainty introduced by stellar jitter. For moderately evolved stars like \thestar{} and \theotherstar{}, radial velocity jitter on relevant timescales can reach $\gtrsim$10 m s$^{-1}$ (see G16 and \S 4.2 for more details).}

\end{deluxetable}

\section{Host Star Characteristics}

\subsection{Spectroscopic Analysis}

In order to obtain precise values for the effective temperature and metallicity of the star, we used the software package SpecMatch \citep{petigura2015} and adopted the spectroscopic analysis method described in G16 for both stars. SpecMatch searches a grid of synthetic model spectra from \citet{coelho2005} to find the best-fit values for \teff , \logg , \feh , mass and radius of the star. We report the effective temperature \teff\ and metallicity \feh\ from the SpecMatch analysis here. We also note that the \logg$_{\mathrm{spec}} = 3.19 \pm 0.07$\ value from the spectroscopic analysis is fully consistent with the asteroseismic determination of \logg$_{\mathrm{AS}} = 3.26 \pm 0.008$ (see next Section for details), so no iteration was needed to recalculate \teff\ and metallicity once asteroseismic parameters had been determined.


\subsection{Asteroseismology}

\begin{figure*}
\begin{center}
\plotone{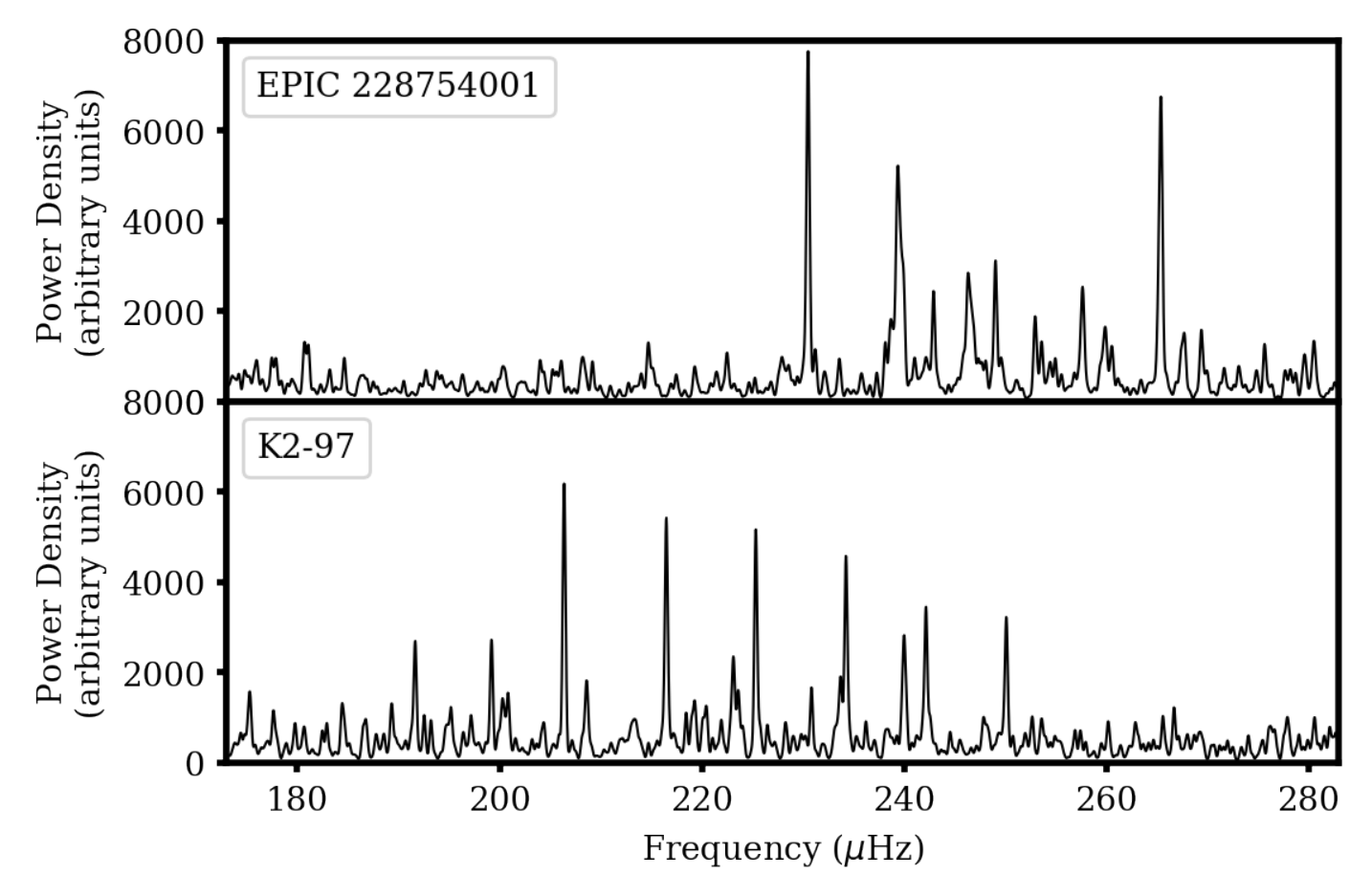}
\caption{Power density of \theotherstar{} (top) and \thestar{} (bottom) estimated from \emph{K2} lightcurves, centered on the frequency range where stellar oscillations can be detected for low luminosity red giant branch (LLRGB) stars. In both cases, stellar oscillations are clearly visible. Note that the power excess of \theotherstar{} does not display a typical Gaussian solar-like oscillation profile due to its proximity to the \emph{K2} long-cadence Nyquist frequency (283 $\mu$Hz).}

\label{seismo}
\end{center}
\end{figure*}

Stellar oscillations are stochastically excited and damped at characteristic frequencies due to turbulence from convection in the outer layers of the star. The characteristic oscillation timescales or frequencies are determined by the internal structure of the star. By measuring the peak frequency of power excess (\numax) and frequency spacing between individual radial orders of oscillation (\dnu), the stellar mass, radius, and density can all be determined to 10\% precision or better.

Similar to G16, we employed asteroseismology using \emph{K2} long-cadence data by measuring stellar oscillation frequencies to determine precise fundamental properties of the evolved host star \theotherstar{}. Figure \ref{seismo} compares the power spectra of \thestar{} and \theotherstar{}. Compared to the power excess of \thestar{} near $\approx$\,220\,\muHz ~(75 minutes), \theotherstar{} oscillates with higher frequencies near $\approx$\,250\,\muHz ~(65 minutes), indicative of a smaller, less evolved RGB star. 

Figure \ref{seismo} also shows that the power excess of \theotherstar{} is less broad and triangular than \thestar{}. This is most likely due to the proximity of the power excess to the long-cadence Nyquist frequency (283.24 $\mu$Hz), causing an attenuation of the oscillation amplitude due to aliasing effects. The proximity to the Nyquist frequency also implies that the real power excess could lie either below or above the Nyquist frequency \citep{chaplin2014, yu2016}. To discern between these scenarios, we applied the method of \citet{yu2016} to distinguish the real power excess from its aliased counterpart. Based on the power-law relation determined by \citet{yu2016}, \dnu\ = 0.262 $\times$ 0.770\numax, as well as a consistent measurement of \dnu\ = 18.46 $\pm$ 0.26 $\mu$Hz both above and below the Nyquist frequency, we find \numax\ = 245.65 $\pm$ 3.51$\mu$Hz, suggesting the true oscillations lie below the Nyquist frequency. To validate this conclusion, we also constructed the global oscillation pattern via the $\varepsilon$-\dnu\ relation \citep{stello2016} for the given \dnu\ value and found the power excess below the Nyquist frequency demonstrates the expected frequency phase shift $\varepsilon$ and matches the expected frequency pattern more precisely. The collapsed \'{e}chelle diagram generated from the \citet{huber2009} pipeline indicates the total power of the $l$ = 2 modes is smaller than that for the $l$ = 0 modes, which also suggests the real power excess is below the Nyquist frequency \citep{yu2016}. Independent asteroseismic analyses using both a separate pipeline for asteroseismic value estimation as well as using lightcurves detrended using different methods recovered asteroseismic parameters in good agreement with the values shown here \citep{north2017}. In addition, the asteroseismic analyses of G16 also strongly agree with our results for \thestar{}. 


\begin{deluxetable*}{lccr}
\tabletypesize{\scriptsize}
\tablecaption{Stellar and Planetary Properties for \thestar{} and \theotherstar{} \label{tbl-star}}
\tablewidth{0pt}
\tablehead{
\colhead{Property} & \colhead{\thestar{}} & \colhead{\theotherstar{}} & \colhead{Source}
}
\startdata
\emph{Kepler} Magnitude & 12.41 & 11.65 & \citet{huber2016} \\
Temperature T$_{\mathrm{eff}}$ & 4790 $\pm$ 90 K & 4840 $\pm$ 90 K & spectroscopy\\
Metallicity [Fe/H] & \fehval{} $\pm$ 0.08 & -0.01 $\pm$ 0.08 & spectroscopy\\
Stellar Mass, $M_{\mathrm{star}}$ & \starmass & \otherstarmass & asteroseismology \\
Stellar Radius, $R_{\mathrm{star}}$ & \starrad & \otherstarrad &  asteroseismology \\
Density, $\rho_{*}$ & 0.0222 $\pm$ 0.0004 g cm$^{-3}$ & 0.0264 $\pm$ 0.0008 g cm$^{-3}$ & asteroseismology\\
log $g$ & 3.26 $\pm$ 0.01 & 3.297 $\pm$ 0.007 & asteroseismology \\
Age & 7.6 $^{+5.5}_{-2.3}$ Gyr & 8.5 $^{+4.5}_{-2.8}$ Gyr & isochrones\\
\hline
Planet Radius, R$_{\mathrm{p}}$ & \planrad & \otherplanrad & GP+transit model \\
Orbital Period $P_{\mathrm{orb}}$ & 8.4061 $\pm$ 0.0015 days & 9.1751 $\pm$ 0.0025 days & GP+transit model\\ 
Planet Mass, M$_{\mathrm{p}}$ &  \planmass & \otherplanmass & RV model \\

\enddata
\tablecomments{All values for the \thestar{} system have been taken from G16, with the exception of the system age, which was recalculated for this publication. See \S 5.1 for a discussion of the system age calculations.}
\end{deluxetable*}

To estimate stellar properties from \numax\ and \dnu, we use the asteroseismic scaling relations of \citet{brown91,kb95}:

\begin{equation}
\frac{\Delta \nu}{\Delta \nu_{\odot}} \approx f_{\Delta \nu} \left(\frac{\rho}{\rho_{\odot}}\right)^{0.5} \: ,
\end{equation}

\begin{equation}
\frac{\nu_{\rm max}}{\nu_{\rm max, \odot}} \approx \frac{g}{{\rm g}_{\odot}} \left(\frac{T_{\rm eff}}{T_{\rm eff, \odot}}\right)^{-0.5} \: .
\end{equation}

\noindent
Equations (1) and (2) can be rearranged to solve for mass and radius:

\begin{equation}
\frac{\rm M}{\rm {\rm M}_\odot}   \approx   \left(\frac{\nu_{\rm
max}}{\nu_{\rm max,
\odot}}\right)^{3}\left(\frac{\Delta \nu}{f_{\Delta \nu}
\Delta \nu_{\odot}}\right)^{-4}\left(\frac{T_{\rm eff}}{T_{\rm
eff, \odot}}\right)^{1.5}   
\end{equation}

\begin{equation}
\frac{\rm R}{\rm R_\odot}   \approx  \left(\frac{\nu_{\rm
max}}{\nu_{\rm max, \odot}}\right)\left(\frac{\Delta
\nu}{f_{\Delta \nu} \Delta \nu_{\odot}}\right)^{-2}\left(\frac{T_{\rm
eff}}{T_{\rm eff, \odot}}\right)^{0.5}.
\end{equation}

Our adopted solar reference values are $\nu_{\rm max, \odot}=3090\,\muHz$ and $\Delta \nu_{\odot}=135.1\,\muHz$ \citep{huber11}, as well as $T_{\rm eff, \odot}=5777$\,K. 

It has been shown that asteroseismically-determined masses are systematically larger than masses determined using other methods, particularly for the most evolved stars \citep{sharma16}. To address this, we also adopt a correction factor of $f_{\Delta \nu}=0.994$ for \thestar{} from G16 and calculate a correction factor $f_{\Delta \nu}=0.998$ for \theotherstar{} following the procedure of \citet{sharma16}. Our final adopted values for the stellar radius, mass, \logg\ and densities of \thestar{} and \theotherstar{} are calculated using these modified asteroseismic scaling relations, and are listed in Table \ref{tbl-star}.

\section{Lightcurve Analysis and Planetary Parameters}

\subsection{Gaussian process transit models}

\begin{figure*}[ht!]
\epsscale{1.0}
\plotone{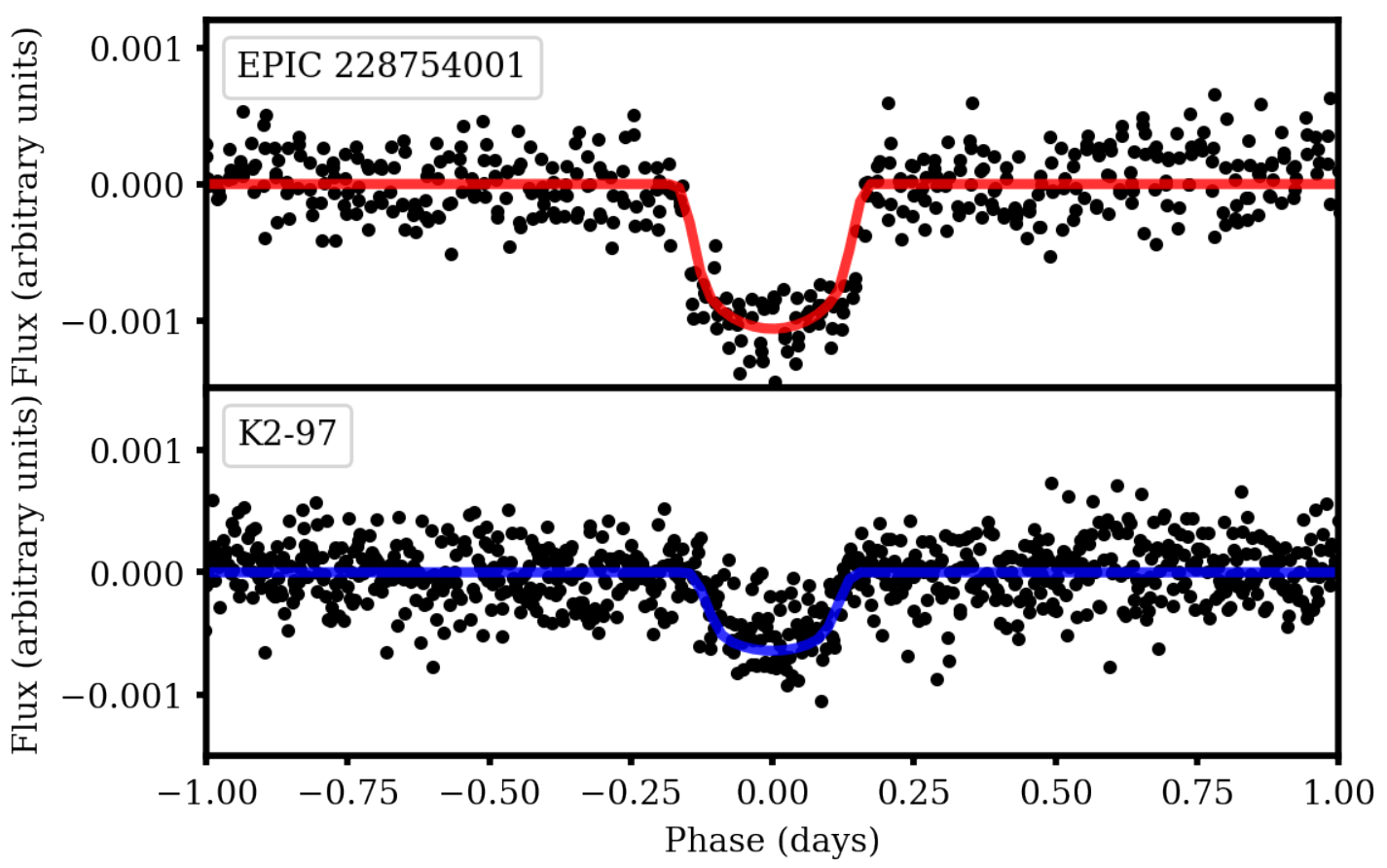}
\caption{Detrended \emph{K2} lightcurves of \theotherstar{} (top) and \thestar{} (bottom), folded at the observed transit period. Preliminary transit fit parameters were established through a box least squares search \citep{kovacs2002}; our final pure transit models \citep{mandel2002} are shown as solid lines.  \label{foldtrans}}
\end{figure*}

\begin{figure}[ht!]
\epsscale{1.22}
\plotone{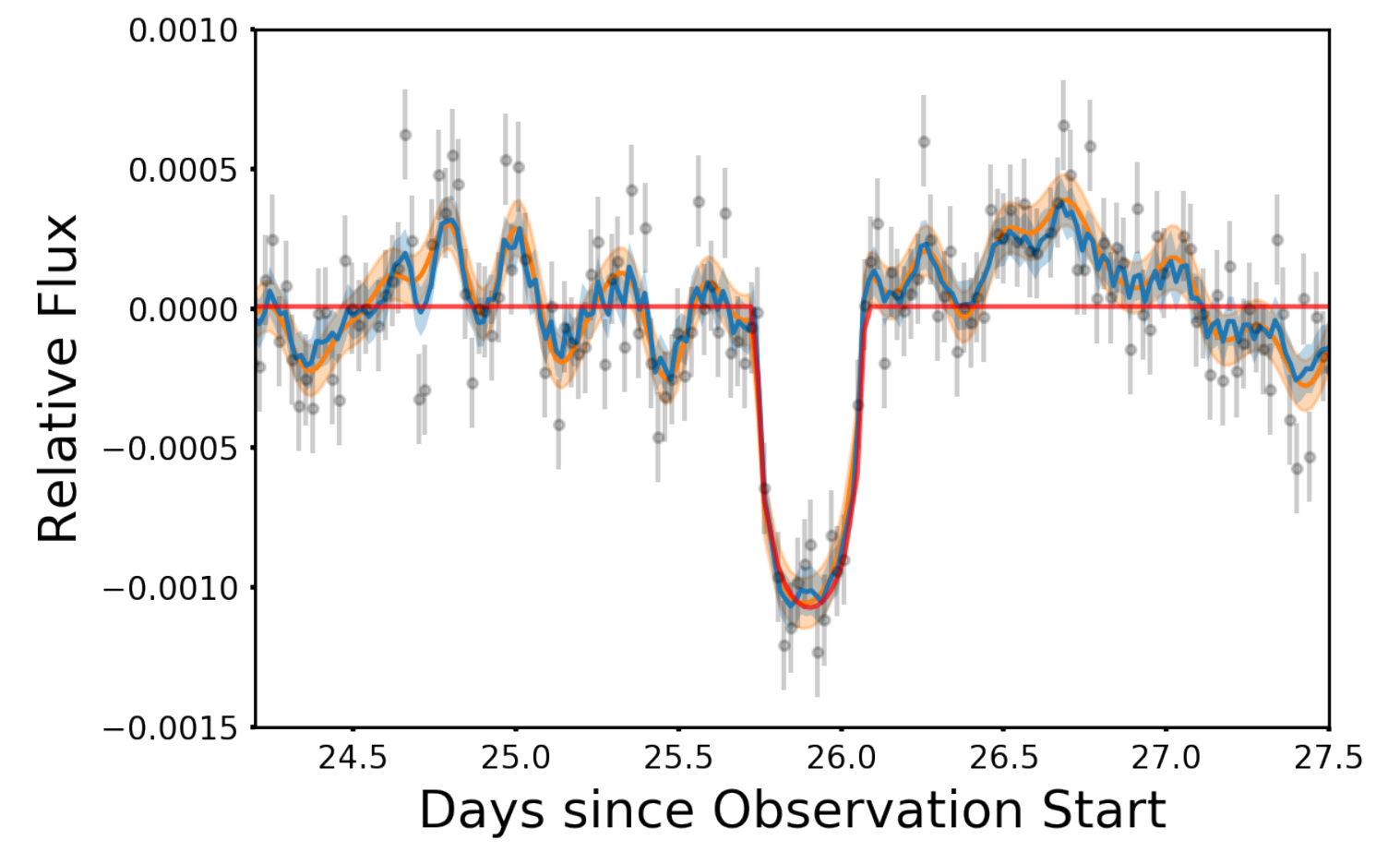}
\caption{Illustration of a transit in the \theotherstar{} lightcurve. The best-fit transit model is shown in red. A combined best-fit transit + squared exponential Gaussian process (SE GP) model is shown in orange, with 1-$\sigma$ model uncertainties shown by the orange shaded region. A combined best-fit transit + simple harmonic oscillator Gaussian process (SHO GP) model is shown with 1-$\sigma$ uncertainties in blue. In addition to having a smaller uncertainties than the SE GP model, the SHO GP model also captures variations on different timescales more accurately, and is physically motivated by the oscillation signal of the star.
\label{GPandtransit}}
\end{figure}


\begin{figure}[ht!]
\epsscale{1.15}
\plotone{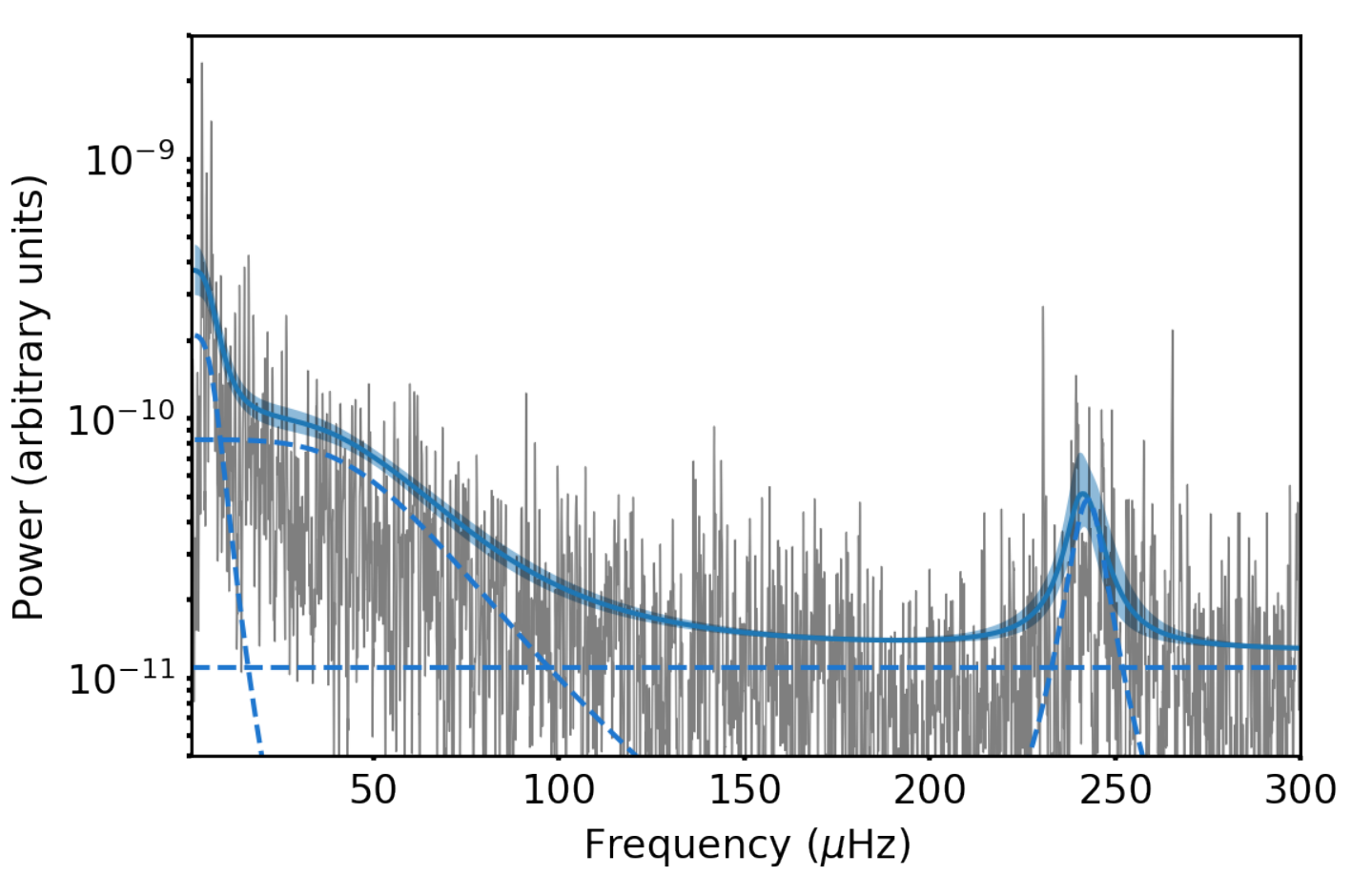}
\caption{The power spectrum of the \theotherstar{} lightcurve (gray) overlaid with the simple harmonic oscillator Gaussian process model (solid blue line). Uncertainties in the model are given by the blue contours. The individual component terms of the Gaussian process model are shown by dotted lines. The two low $Q$ components account for the granulation noise signal at low frequencies. The high $Q$ component traces the envelope of stellar oscillation signal and allows us to estimate the frequency of maximum power of the stellar oscillations, and thus determine \numax\ from the time domain. 
\label{SHOGPpowspec}}
\end{figure}

\begin{figure}[ht!]
\epsscale{1.0}
\plotone{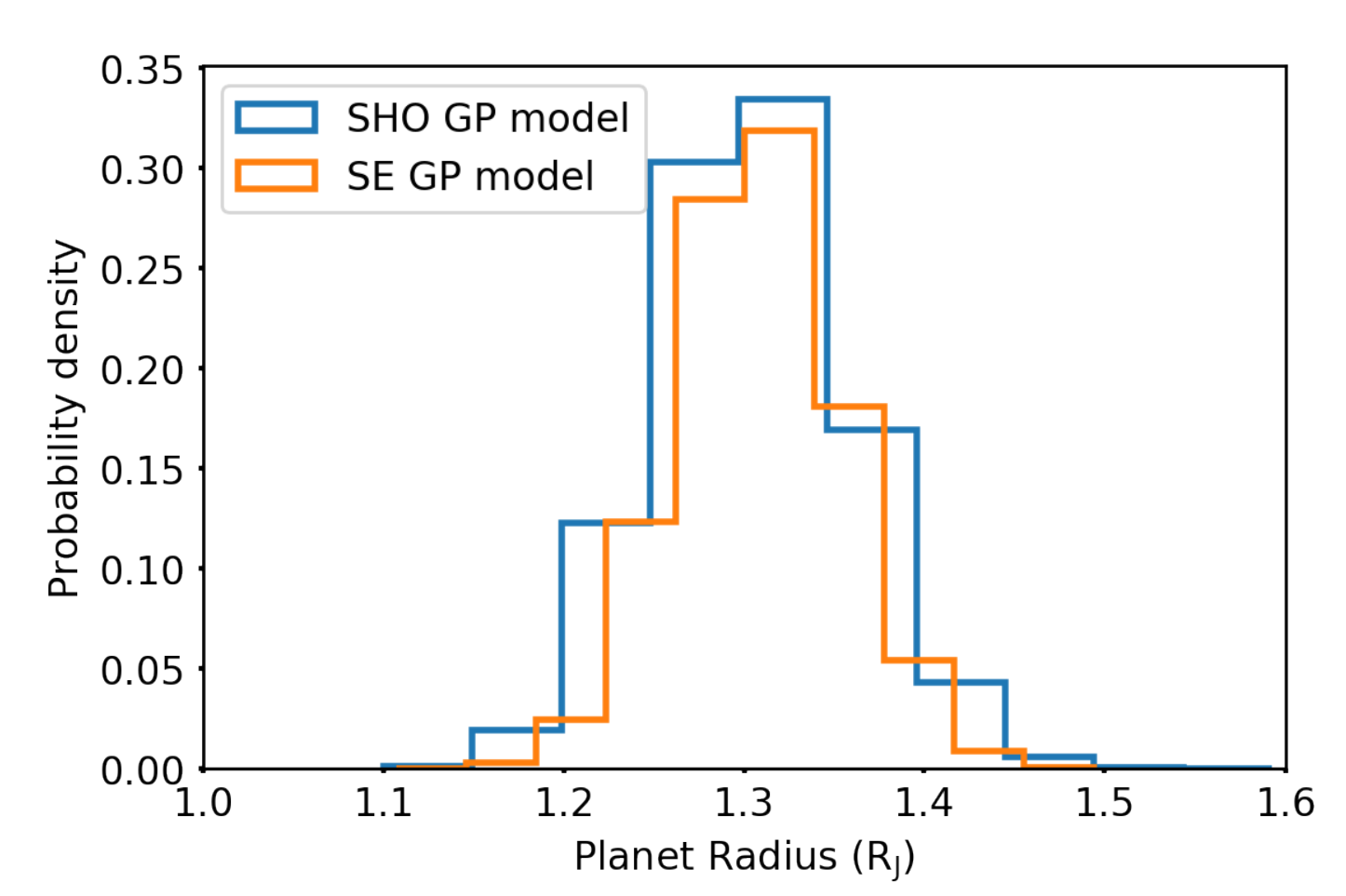}
\caption{Posterior distributions of planet radius based on our stellar parameters derived from asteroseismology and transit depth measured in our transit + squared exponential Gaussian process model (SE GP model, orange) and our transit + simple harmonic oscillator Gaussian process model (SHO GP model, blue) for \theotherstar{.01}. Parameters differ between the two models, but both provide estimates of $R_p/R_*$ which can be converted into planet radius and directly compared. We find that our squared exponential (SE) GP model strongly agrees with our simple harmonic oscillator (SHO) GP model.
\label{planradcomp}}
\end{figure}

\begin{figure*}[ht!]
\epsscale{1.0}
\plotone{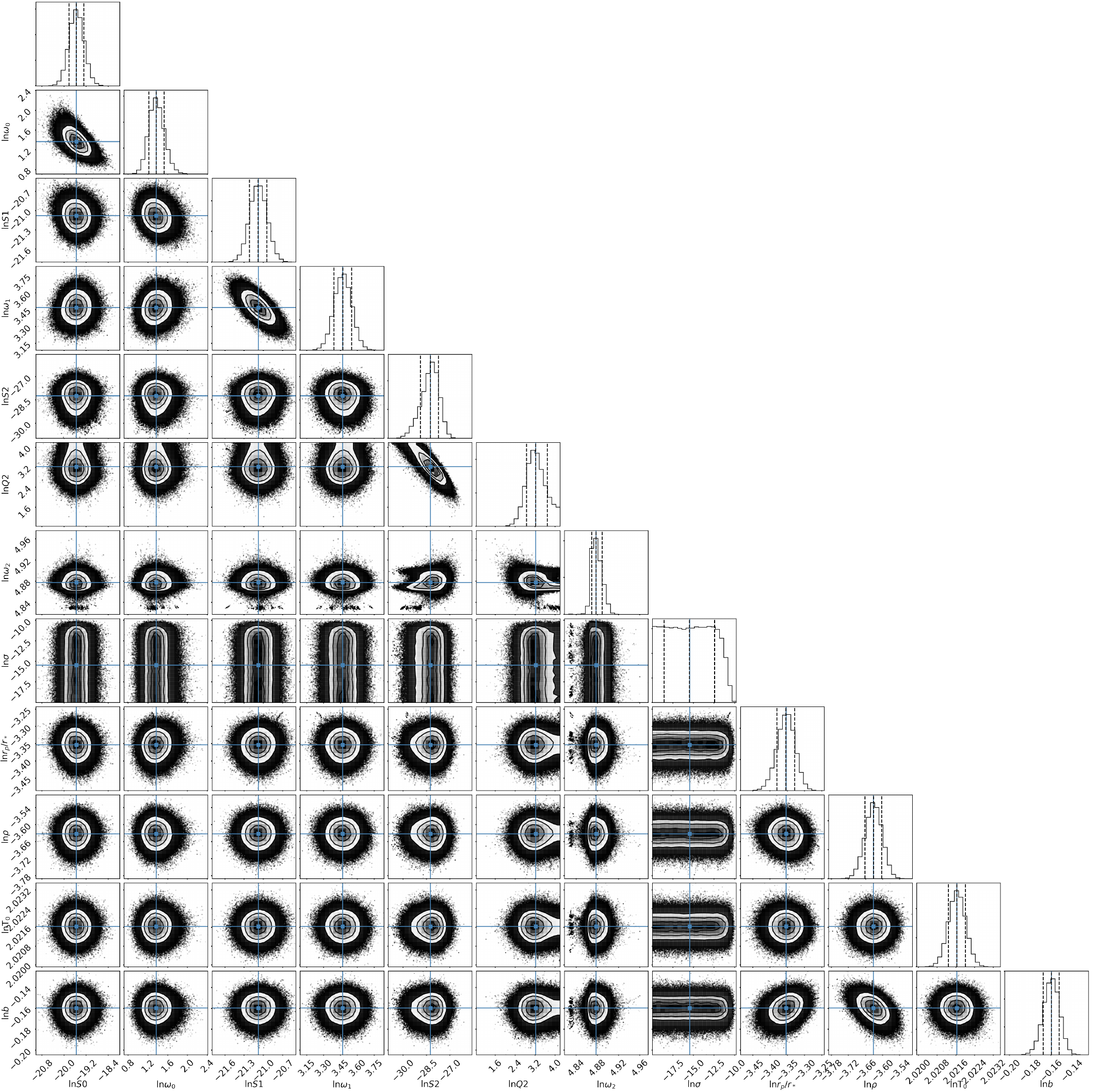}
\caption{Posterior distributions for the complete transit + GP model of \theotherstar{}. The first 8 parameters are part of the GP model, whereas the last 4 are components of the transit model. Individual parameter posterior distributions are shown along the diagonal, while correlations between two parameters are shown as the off-diagonal, two-dimensional distributions. Median values are indicated by the blue lines; dotted lines indicate 1-$\sigma$ uncertainties. Priors are discussed in further detail within the text.
\label{triangle}}
\end{figure*}


The transits of \thestar{b} and \theotherstar{.01} were first identified using the box least-squares procedure described in G16 and \S 2.1 \citep{kovacs2002}. The detrended lightcurves, phase folded at the period detected by the box least-squares search and fit with best-fit transit models, are shown in Figure \ref{foldtrans}.

Evolved stars display correlated stellar variation on timescales of hours to weeks due to stellar granulation and oscillation \citep{mathur2012}, leading to systematic errors in transit parameter estimation \citep{carter2009, barclay2015}. Thus, a stochastically-driven and damped simple harmonic oscillator can be used to both describe the stellar oscillation and granulation noise in a lightcurve as well as characterize the fundamental physical properties of the star.

In G16, we used a squared exponential Gaussian process estimation model to remove stellar variability in the K2 lightcurve and measure the transit depth of \thestar{b} precisely. Here, we used a Gaussian process estimation kernel that assumes stellar variability can be described by a stochastically-driven damped simple harmonic oscillator, modified from the method of G16. We also present results using the previously tested squared exponential Gaussian process kernel, which has been successfully applied to remove correlated noise in various one dimensional datasets in the past \citep{gibson2012,dawson2014,haywood2014,barclay2015,grunblatt2015, grunblatt2016}. 


We describe the covariance of the time-series data as an N$\times$N matrix $\mathbf{\Sigma}$ where 
\begin{equation}
\Sigma_{ij} = \sigma_i^2\delta_{ij} + k(\tau_{ij}) 
\end{equation}
where $\sigma_i$ is the observational uncertainty, $\delta_{ij}$ is the Kronecker delta, and $k (\tau_{ij})$ is the so-called covariance kernel function that quantifies the correlations between times $t_i$ and $t_j$ \citep{rasmussen2006}. 

Following \citet{foreman-mackey2017}, the kernel function we use can be expressed as 
\begin{multline}
k(\tau_{ij}) = \sum\limits_{n=1}^N    [a_n \mathrm{exp} (-c_n \tau_{ij}) \mathrm{cos} (d_n \tau_{ij}) \\
+ b_n \mathrm{exp} (-c_n \tau_{ij}) \mathrm{cos} (d_n \tau_{ij}) ]
\end{multline}
 
 where $a_n$, $b_n$, $c_n$ and $d_n$ are a set of constants that define the $n$th term in our kernel function. We then redefine these constants $a_n$, $b_n$, $c_n$ and $d_n$ as simple harmonic oscillator components $Q_n$, $\omega_{0,n}$ and $S_{0,n}$ such that
\begin{multline}
k(\tau_{ij}) = S_0 \omega_0 Q e^{-\frac{\omega_0 \tau_{ij}}{2Q}} \times \\
\begin{cases} \mathrm{cosh}( \eta \omega_0 \tau_{ij}) + \frac{1}{2\eta Q}\mathrm{sinh} ( \eta \omega_0 \tau_{ij}), & 0 < Q < 1/2 \\
2 (1 + \omega_0 \tau_{ij}), & Q = 1/2 \\
\mathrm{cos}(\eta \omega_0 \tau_{ij}) + \frac{1}{2\eta Q}\mathrm{sin} (\eta \omega_0 \tau_{ij}), & 1/2 < Q, \end{cases}
\end{multline}
where $Q_n$ represents the quality factor or damping coefficient of the $n$th simple harmonic oscillator, $\omega_{0,n}$ represents the resonant frequency of the $n$th simple harmonic oscillator, $S_{0,n}$ is proportional to the power at $\omega$ = $\omega_{0,n}$, and $\eta = \sqrt{1 - (4Q^2)^{-1}}$. We find that we can describe the stellar variability seen in our data as a sum of three simple harmonic oscillator components, similar to many asteroseismic models used to describe stellar oscillations \citep[eg.,][]{huber2009}. This allows us to create a physically motivated model of stellar variability from which we can produce rigorous probabilistic measurements of asteroseismic quantities using only time domain information. 

Our simple harmonic oscillator Gaussian process model consists of three main components: two $Q = 1/\sqrt{2}$ terms, which are commonly used to model granulation in asteroseismic analyses \citep{harvey1985, huber2009, kallinger2014}, and one $Q \gg 1$ term, which has been shown to describe stellar oscillations effectively \citep{foreman-mackey2017}, to describe the envelope of stellar oscillation signal. The resonant frequency $\omega_0$ of this component of is thus an independent estimate of \numax, and we compare our asteroseismic \numax\ measurement made from analysis in the frequency domain to the \numax\ we generate here through a pure time domain analysis. We find good agreement between our independent estimates of \numax\ for \theotherstar{} using both traditional asteroseismic analysis methods (\numax\ = 245.65 $\pm$ 3.51 $\mu$Hz) and our simple harmonic oscillator Gaussian process model estimate ($\nu_{\mathrm{max, GP}}$ = 241.8 $\pm$ 1.9 $\mu$Hz).

Following the procedure of G16, we incorporate a transit model with initial parameters determined by the box least-squares analysis as the mean function from which residuals and the Gaussian process kernel parameters are estimated. By exploring probability space through an MCMC routine where a likelihood for the combined transit and variability model is calculated repeatedly, we simultaneously optimize both the stellar variability and transit parameters. The logarithm of the posterior likelihood of our model is given as
\begin{equation} \label{eq:2}
\mathrm{log}[\mathcal{L}(\mathbf{r})] = -\frac{1}{2}\mathbf{r}^\mathrm{T}\mathbf{\Sigma}^{-1}\mathbf{r} - \frac{1}{2} \mathrm{log} |\mathbf{\Sigma}| - \frac{n}{2} \mathrm{log}(2\pi),
\end{equation}
where $\mathbf{r}$ is the vector of residuals of the data after removal of the mean function (in our case, $\mathbf{r}$ is the lightcurve signal minus the transit model), and $n$ the number of data points. 

We repeat this process using both the new simple harmonic oscillator Gaussian process estimator as well as the squared exponential Gaussian process estimator. We illustrate our transit + GP models and uncertainties in the time domain in Figure \ref{GPandtransit}, as well as our simple harmonic oscillator GP model in the frequency domain in Figure \ref{SHOGPpowspec}. We find that our simple harmonic oscillator Gaussian process estimation is able to capture variation on a wider range of timescales than the squared exponential Gaussian process estimation, and also features smaller uncertainty distributions in the time domain. In addition, the simple harmonic oscillator model exploits the tridiagonal structure of a covariance matrix generated by a mixture of exponentials such that it scales linearly, rather than cubicly, with the size of the input dataset. This means the squared exponential Gaussian process estimation takes over an order of magnitude more time to generate for the entire lightcurve than the simple harmonic oscillator model despite having less than half the number of parameters. Furthermore, the squared exponential estimate provides a poor estimate of the appearance of the data in the frequency domain, whereas the simple harmonic oscillator estimate is able to reproduce both an estimate of the granulation background as well as the stellar oscillation signal, two of the strongest features of the stellar signal in the frequency domain. The similarity between the simple harmonic oscillator estimate and the power spectral density estimate from the lightcurve is particularly remarkable considering all fitting was done using time domain information, suggesting that this simple harmonic oscillator estimation technique may be a valuable prototype for designing a technique to perform ensemble asteroseismology using only time domain information \citep{brewer2009, foreman-mackey2017}.

Due to the benefits from employing the simple harmonic oscillator Gaussian process estimation technique to extract the planet to star radius ratio, we choose to use the results from this model as our accepted values for calculating planet radius. We show the best-fit results for selected parameters of interest in Table \ref{tbl-2}. The posterior distributions of the planet radius estimated with both methods are shown in Figure \ref{planradcomp}, illustrating that planet radius estimates by both Gaussian process techniques are in very good agreement.

Figure \ref{triangle} illustrates the parameter distributions for the full transit+GP model. All parameters are sampled in logarithmic space. The first nine parameters are simple harmonic oscillator components terms of the model, as well as the white noise $\sigma$. The last four parameters of the model are transit parameters $R_p/R_*$, stellar density $\rho$, phase parameter $T_0$, and impact parameter $b$, . Correlations between $b$ and $R_p/R_*$ can be seen. Uniform box priors were placed on all GP parameters to ensure physical values. In addition, ln$\omega_{0,0}$ has a strict lower bound of 1.1 as the data quality at frequencies lower than 3 $\mu$Hz is too poor to warrant modeling. ln$Q_2$ has a strict upper bound of 4.2 to ensure that the envelope of stellar oscillations is modeled as opposed to individual frequencies of stellar oscillation (which correspond to higher $Q$ values), and $\omega_{0,2}$ has bounds of 200 and 280 $\mu$Hz to ensure that the excess modeled corresponds to the  asteroseismic excess determined previously. The lower bound of the white noise parameter ln$\sigma$ posterior distribution is also set by a uniform box prior, as the median absolute deviation of the lightcurve (162 ppm, not a variable in our model) is sufficient to capture the uncorrelated variability in our data and thus any additional white noise below this level is equally likely given this dataset. A Gaussian prior has been placed on $\rho$ according to its asteroseismic determination in \S 3.2. Eccentricity is fixed to zero for our transit model, based on arguments explained in \S 5.3.

In addition, the quadratic limb darkening parameters $\gamma_1$ and $\gamma_2$ in our transit model were fixed to the \citep{claret2011} stellar atmosphere model grid values of 0.6505 and 0.1041, respectively. These values correspond to the stellar model atmosphere closest to the measured temperature, surface gravity, and metallicity of the host star. As \citet{barclay2015} demonstrate that limb darkening parameters are poorly constrained by the transits of a giant planet orbiting a giant star with 4 years of \emph{Kepler} photometry, our much smaller sample of transits, all of which are polluted by stellar variability, would not be sufficient to constrain limb darkening.

In order to evaluate parameter convergence, the Gelman-Rubin statistic was calculated for each parameter distribution and forced to reach 1.01 or smaller \citep{gelman1992}. In order to achieve this, 30 Monte Carlo Markov Chains with 50,000 steps each were used to produce parameter distributions.

\subsection{Radial Velocity Analysis, Planetary Confirmation, and False Positive Assessment}

We modeled the Keck/HIRES RV measurements of \thestar{} and \theotherstar{} following the method of G16, with slight modifications. Similarly to G16, we produced an initial fit for the systems using the publicly available Python package \texttt{RadVel} \citep{radvel}, and then fit the data independently as a Keplerian system with amplitude $K$, phase $\phi$, white noise $\sigma$, and radial velocity zeropoint $z$, and a period $\theta$ predetermined and fixed from the transit analysis. 

We assume the eccentricity of the planet is fixed to zero in our transit and radial velocity analysis based on dynamical arguments presented in \S 5.3. Nevertheless, the data is not sufficient to precisely constrain the eccentricity of this system. \citet{jones2017} explore the possibilities of eccentricity in this system in further detail.

\begin{figure*}[ht!]
\epsscale{1.0}
\plottwo{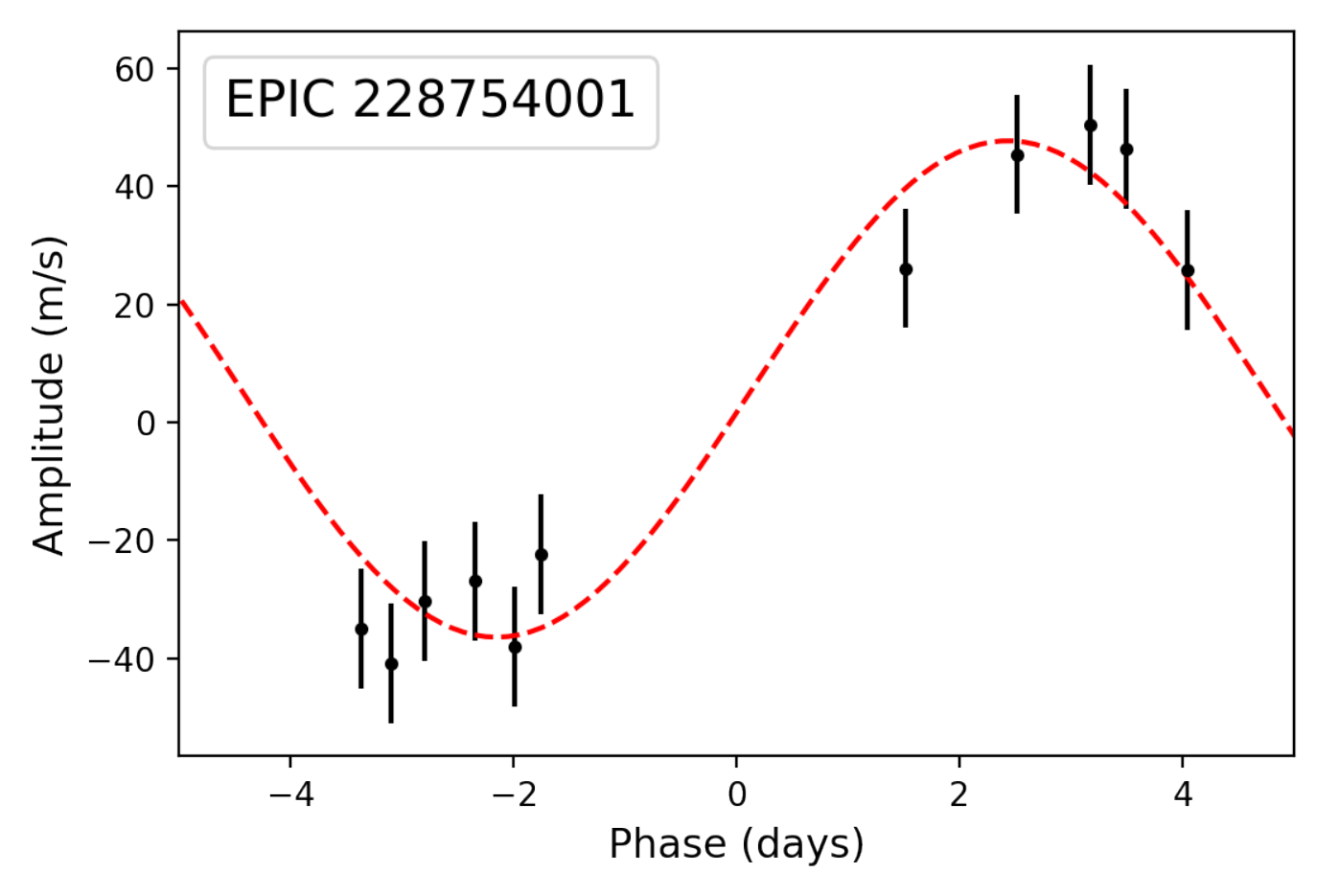}{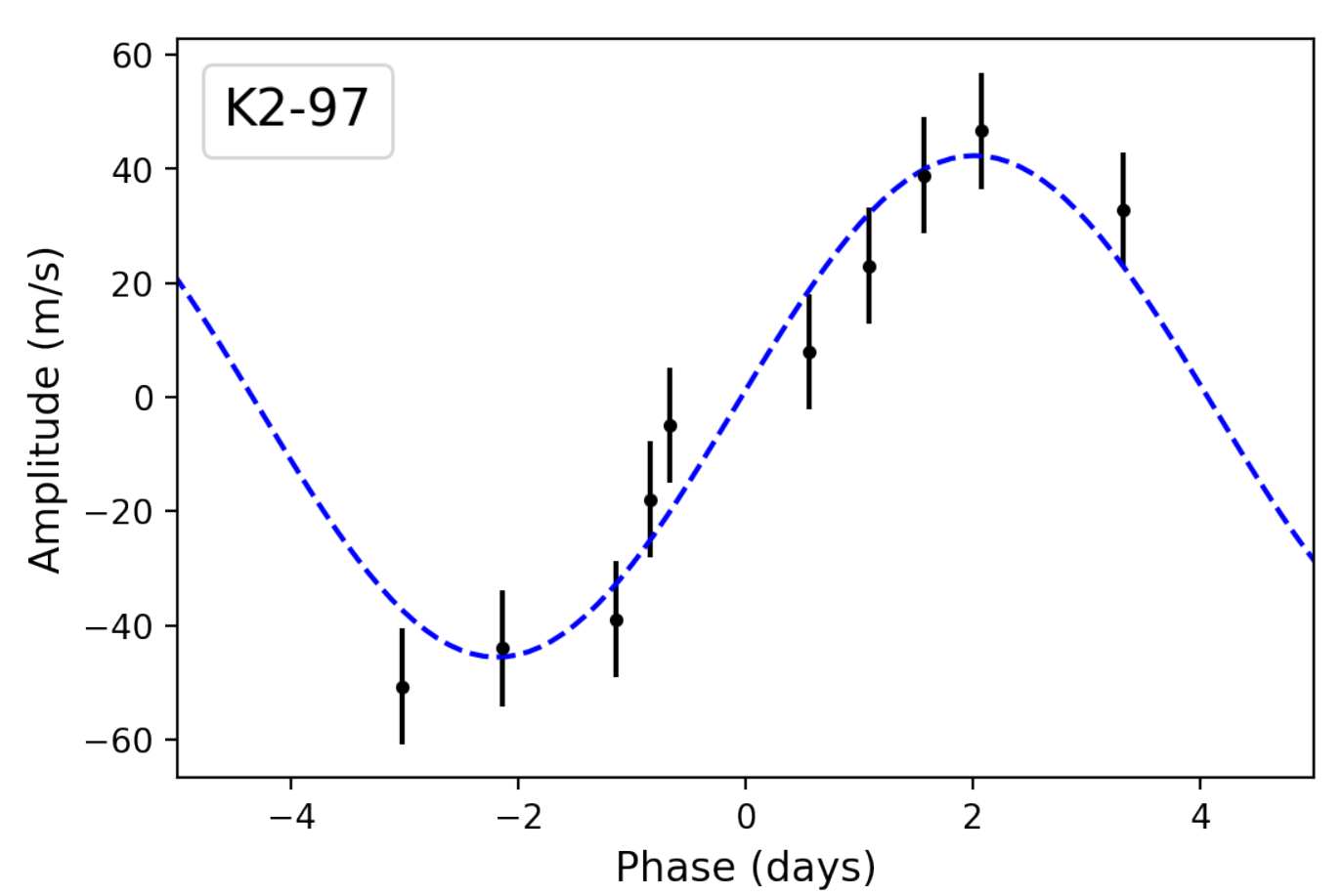}
\caption{Black points show Keck/HIRES radial velocity measurements of the \thestar{b} and \theotherstar{.01} systems, phase-folded at their orbital periods derived from lightcurve analysis. Errors correspond to the measurement errors of the instrument added in quadrature to the measured astrophysical jitter. The dashed colored curves correspond to the one-planet Keplerian orbit fit to the data, using the median value of the posterior distribution for each fitted Keplerian orbital parameter. Parameter posterior distributions were determined through MCMC analysis with $\texttt{emcee}$. \label{rv}}
\end{figure*}

Due to the relatively high degree of scatter within our radial velocity measurements, and the known increase in radial velocity scatter due to stellar jitter as stars evolve up the red giant branch \citep{huber2011}, we fit for the astrophysical white noise error and add it to our radial velocity measurement errors in quadrature, finding typical errors of 10--15 m s$^{-1}$. Non-transiting planets orbiting at different orbital periods may also add additional uncertainty to our measurements. We have probed modestly for these planets by collecting radial velocity measurements spanning multiple orbital periods of the transiting planet in both systems, confirming that the dominant periodic radial velocity signal coincides with the transit events. Median values and uncertainties on Keplerian model parameters were determined using Monte Carlo Markov Chain analysis powered by \texttt{emcee} \citep{foremanmackey2013}. We illustrate the radial velocity measurements of both systems as well as the best-fit Keplerian models in Figure \ref{rv}.

\begin{figure*}[ht!]
\epsscale{1.0}
\plotone{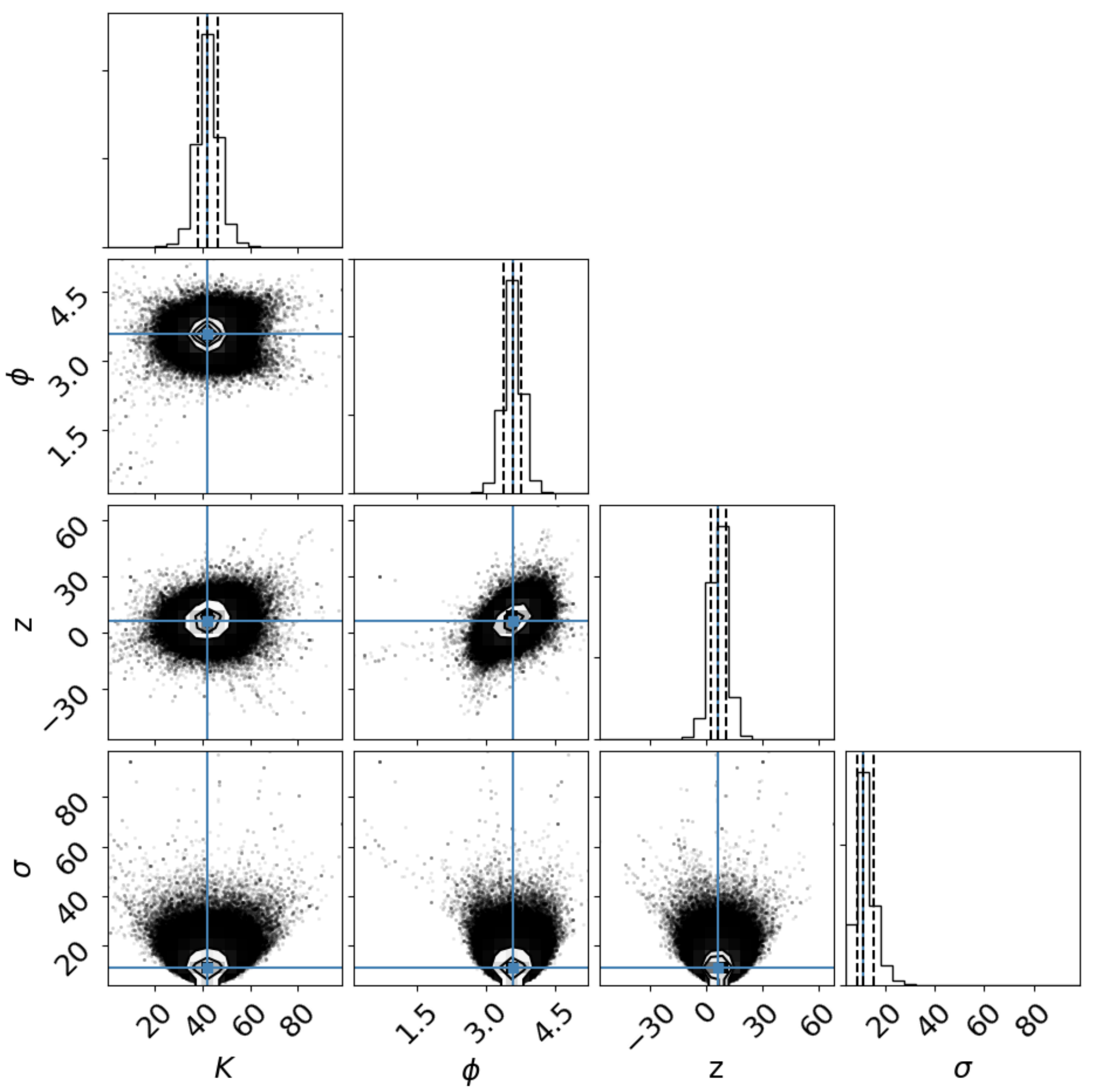}
\caption{Posterior distributions for the complete RV model of \theotherstar{.01}. Individual parameter posterior distributions are shown along the diagonal, while correlations between two parameters are shown as the off-diagonal, two-dimensional distributions. Median values are indicated by the blue lines; dotted lines indicate 1-$\sigma$ uncertainties.
\label{rvtriangle}}
\end{figure*}

Figure \ref{rvtriangle} illustrates the posterior distributions for the RV model amplitude $K$, phase $\phi$, zeropoint $z$, and uncorrelated uncertainty $\sigma$. In order to evaluate parameter convergence, the Gelman-Rubin statistic was calculated for each parameter distribution and forced to reach 1.01 or smaller \citep{gelman1992}. In order to achieve this, 30 Monte Carlo Markov Chains with 50,000 steps each were used to produce parameter distributions.

The initial confirmation of the \thestar{b} system included the three earliest Keck/HIRES measurements shown here as well as radial velocities measured by the Automated Planet Finder (APF) Levy Spectrometer at the Lick Observatory in California. Due to the relatively large uncertainties on the APF measurements, the earlier mass estimates were dominated by the Keck/HIRES data. However, the small number of Keck/HIRES measurements spanned less than 10\% of the entire orbit. This limited coverage, as well as an overly conservative estimate of stellar jitter, resulted in an overestimate of the mass of \thestar{b} in G16. The additional coverage by Keck/HIRES since the publication of G16 has negated the issues brought by the relatively large uncertainties of the APF measurements, and effectively expanded the radial velocity phase coverage to $>$50\%. This revealed that the previous characterization of stellar jitter was an underestimate and the planet mass was significantly lower than estimated in G16.


\begin{deluxetable}{lcccccrrrrrrrrcrl}
\tabletypesize{\scriptsize}
\tablecaption{Posterior Probabilities from Lightcurve and Radial Velocity MCMC Modeling of \theotherstar{}\label{tbl-2}}
\tablewidth{0pt}
\tablehead{
\colhead{Parameter} & \colhead{Posterior Value}  & \colhead{Prior}
}
\startdata
\centering
 $\rho$ (g cm$^{-3}$) & $0.0264^{+0.0008}_{-0.0007}$ & $\mathcal{N}$(0.0264; 0.0008) \\
T$_0$ (BJD-2454833) & $2757.1491^{+0.008}_{-0.009}$ & $\mathcal{U}$(5.5; 9.5) \\
$P_{\mathrm{orb}}$ (days) & $9.1751^{+0.0023}_{-0.0027}$ & $\mathcal{U}$(9.0; 9.4) \\ 
$b$ & $0.848^{+0.007}_{-0.008}$ & $\mathcal{U}$(0.0, 1.0 + $R_p/R_*$)\\
$R_p / R_{*}$ & $0.0325^{+0.0014}_{-0.0011}$ & $\mathcal{U}$(0.0, 0.5) \\
$\numax_{, \mathrm{GP}}$ ($\mu$Hz) & $241.8^{+1.9}_{-1.9}$ & $\mathcal{U}$(120, 280))\\
K (m s$^{-1}$) & $42.1^{+4.3}_{-4.2}$ & \\
T$_{0,\mathrm{RV}}$ (BKJD \% $P_{\mathrm{orb}}$) & $3.57^{+0.19}_{-0.19}$ & $\mathcal{U}$(0.0,
$P_{\mathrm{orb}}$) \\
$\sigma_{\mathrm{RV}}$ (m s$^{-1}$) &  $11.5^{+4.1}_{-2.6}$ & $\mathcal{U}$(0, 100) \\
\enddata 
  
\tablecomments{$\mathcal{N}$ indicates a normal distribution with mean and standard deviation given respectively. $\mathcal{U}$ indicates a uniform distribution between the two given boundaries. Ephemerides were fit relative to the first measurement in the sample and then later converted to Barycentric \emph{Kepler} Julian Date (BKJD).} 

\end{deluxetable}

We quantitatively evaluated false positive scenarios for EPIC~228754001.1 as in G16 and more thoroughly described in \citet{Gaidos2016}, using our adaptive optics (AO) imaging and lack of a long-term trend in our radial velocity measurements of \theotherstar{} to rule out a background eclipsing binaries or hierarchical triple (companion eclipsing binary).  We reject these scenarios because the nearly 8~hr transit duration is much too long compared to that expected for an eclipsing binary with the same period, provided that the system is not highly eccentric ($e$ $>$ 0.3), and our radial velocity measurements rule out a scenario involving two stellar mass objects. Preliminary evidence from our radial velocity data also suggests that an eccentricity of $e$ $>$ 0.3 is unlikely for this system, but a full exploration of eccentricity scenarios is beyond the scope of this article (see \S 5.3 for more details). Furthermore, a background evolved star that was unresolved by our AO imaging is too unlikely $\ll 2 \times 10^{-7}$ and the dilution too high by the foreground (target) star to explain the signal. Evolved companions are ruled out by our AO imaging to within 0.2'' and stellar counterparts within $\sim 1$ AU are ruled out by the absence of an RV drift.

We cannot rule out companions that could cause a small systematic error in planet radius due to dilution of the transit signal.  However, to change the planet radius by one standard error the \emph{minimum} contrast ratio in the \emph{Kepler} bandpass must be 0.1.  If the star is cooler than EPIC~228754001 (likely, since a hotter, more massive star would be more evolved) then the contrast in the $K$-band of our NIRC2 imaging would be even higher.  We can rule out all such stars exterior to 0.15 arcsec ($\sim 50$~AU) of the primary; absence of a significant drift in the Doppler data or a second set of lines in the HIRES spectrum rules out stellar companions within about 1~AU.  Regardless, transit dilution by an unresolved companion would mean that the planet is actually \emph{larger} than we estimate and inflation even more likely.

\begin{figure}[ht!]
\epsscale{1.1}
\plotone{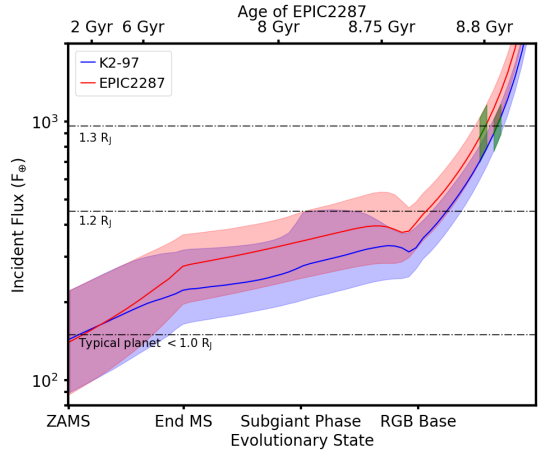}
\caption{Incident flux as a function of evolutionary state for \thestar{b} and \theotherstar{.01}. The current incident flux on the planets is denoted in green. Solid blue and red lines and shaded areas show the median and 1-$\sigma$ confidence interval considering uncertainties in stellar mass and metallicity. The black dashed lines correspond to the median incident fluxes for known populations of hot gas giant planets of different radii \citep[][NASA Exoplanet Archive, 9/14/2017]{demory2011}. The top axis shows representative ages for the best-fit stellar parameters of \theotherstar{}.\label{evol}}
\end{figure}

\begin{figure*}[ht!]
\epsscale{1.0}
\plottwo{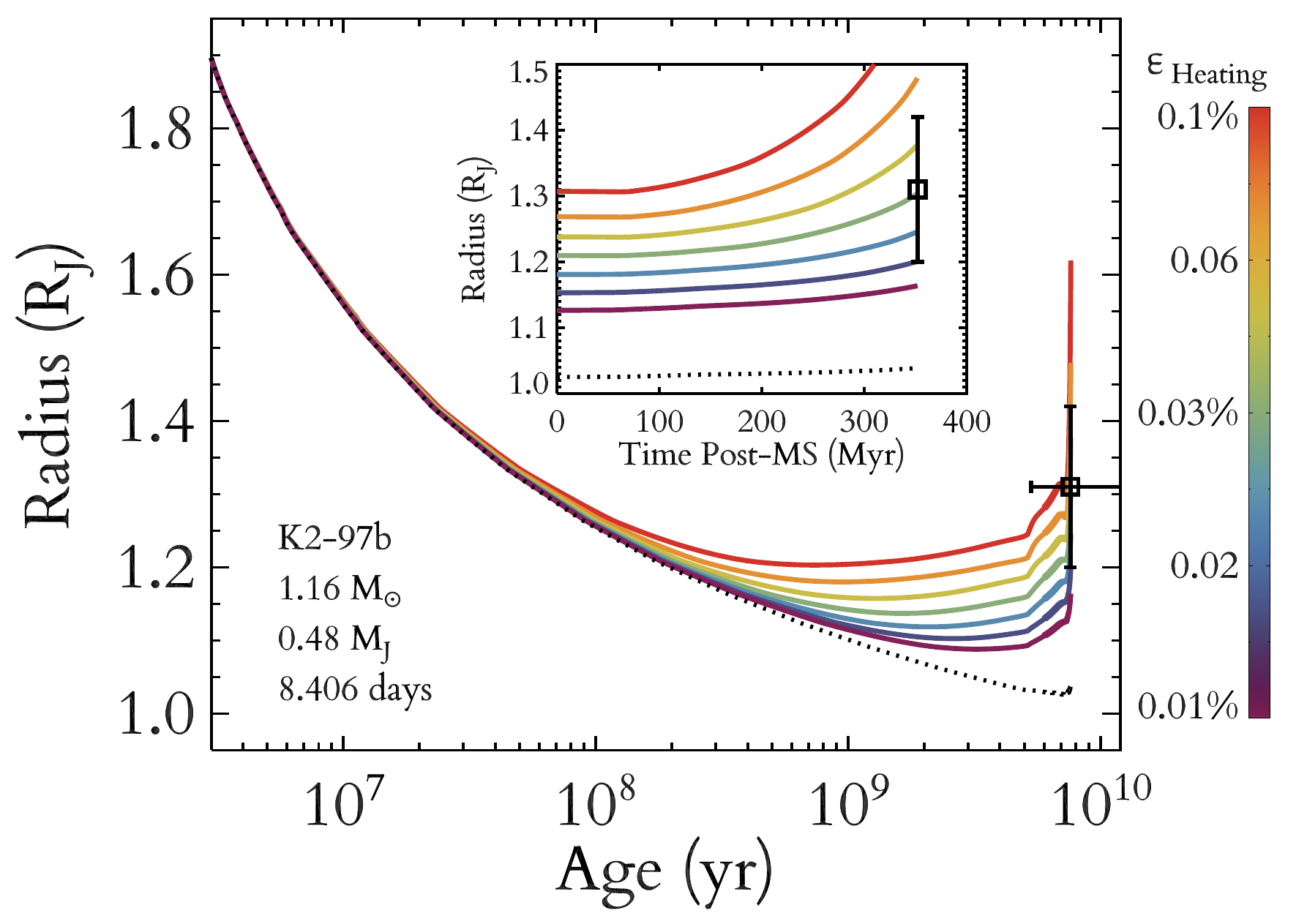}{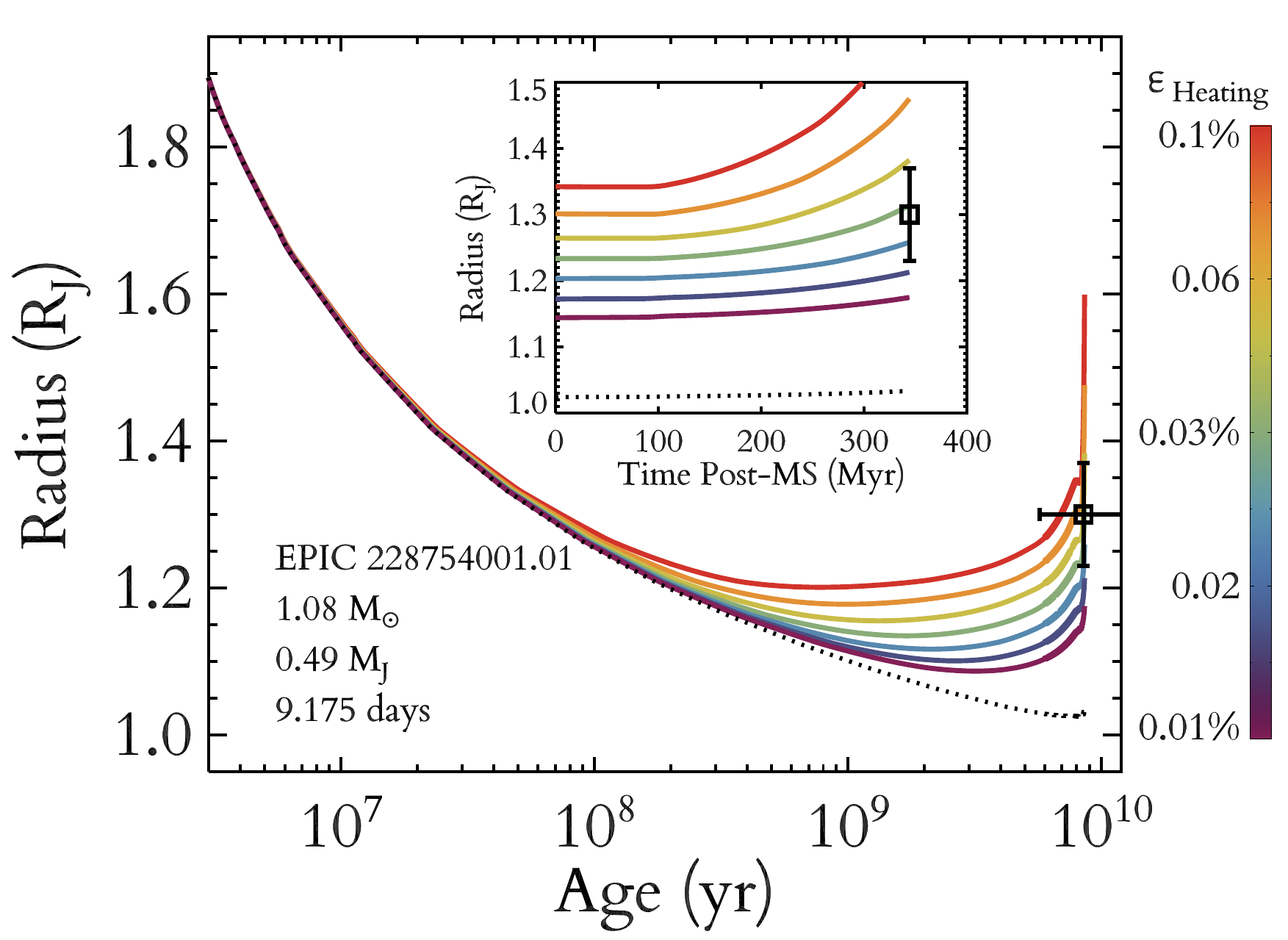}
\caption{Planetary radius as a function of time for \thestar{b} (left) and \theotherstar{.01} (right), shown for various different values of heating efficiency. We assume the best-fit values for the stellar mass and the planetary mass and radius, and a planetary composition of a H/He envelope surrounding a 10 M$_\oplus$ core of heavier elements. The dotted line corresponds to a scenario with no planetary heating. The inset shows the post-main sequence evolution at a finer time resolution. The measured planet radii are consistent with a heating efficiency of 0.03\%$^{+0.04\%}_{-0.02\%}$ and  0.03\%$^{+0.3\%}_{-0.1\%}$, respectively.
\label{heating}}
\end{figure*}

\begin{figure}[ht!]
\epsscale{1.0}
\plotone{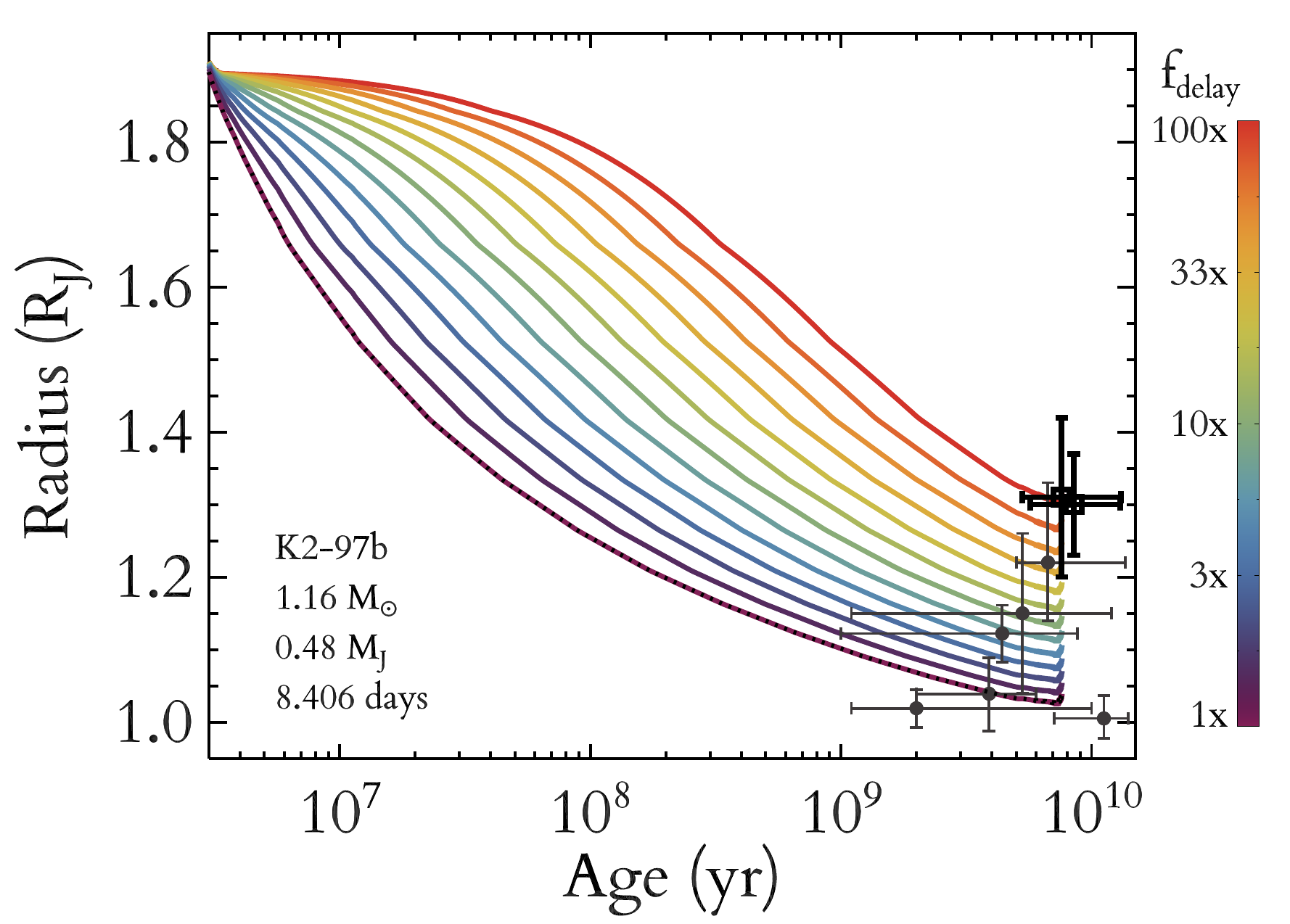}
\caption{Planetary radius as a function of time for \thestar{b} and \theotherstar{.01} (bold), as well other similar mass planets with similar main sequence fluxes orbiting main sequence stars. Colored tracks represent scenarios where planets begin at an initial radius of 1.85 R$_\mathrm{J}$ and then contract according to the Kelvin-Helmholtz timescale delayed by the factor given by the color of the track. All main sequence planets seem to lie on tracks that would favor different delayed cooling factors than the post-main sequence planets studied here.
\label{cooling}}
\end{figure}

\section{Constraining Planet Inflation Scenarios}

\subsection{Irradiation Histories of \thestar{b} and \theotherstar{.01}}

Planets with orbital periods of $<$30 days will experience levels of irradiation comparable to typical hot Jupiters for more than 100 Myr during post-main sequence evolution. Thus, we can test planet inflation mechanisms by examining how planets respond to increasing irradiation as the host star leaves the main sequence. Following the nomenclature of \citet{lopez2016}, if the inflation mechanism requires direct heating and thus falls into Class I, the planet's radius should increase around a post-main sequence star. However, if the inflation mechanism falls into Class II, requiring delayed cooling, there should be no effect on planet radius as a star enters the red giant phase, and re-inflation will not occur.  As \thestar{b} and \theotherstar{.01} are inflated now but  may not have received irradiation significantly above the inflation threshold on the main sequence, they provide valuable tests for the re-inflation hypothesis. Furthermore, these systems can be used to constrain the mechanisms of heat transfer and dissipation within planets \citep[e.g.,][]{tremblin2017}.

To trace the incident flux history of both planets we used a grid of Parsec v2.1 evolutionary tracks \citep{bressan12} with metallicities ranging from $\feh = -0.18$ to $0.6$\,dex and masses ranging from $0.8-1.8$\msun. Compared to G16, we used an improved Monte-Carlo sampling scheme by interpolating evolutionary tracks to a given mass and metallicity following normal distributions with the values given in Table \ref{tbl-star}, and tracing the incident flux across equal evolutionary states as indicated by the ``phase'' parameter in Parsec models. We performed 1000 iterations for each system, and the resulting probability distributions are shown as a function of evolutionary state in Figure \ref{evol}. We note that each evolutionary state corresponds to a different age depending on stellar mass and metallicity. Representative ages for the best-fit stellar parameters of \theotherstar{} are given on the upper x-axis. Current incident flux and age ranges for the planets were determined by restricting models to within 1-$\sigma$ of the measured temperature and radius of each system (Table \ref{tbl-star}).


Figure \ref{evol} demonstrates that both planets lie near the \citet{demory2011} empirical threshold for inflated planets at the zero age main sequence. Planets below this threshold have typical planet radii below 1.0 R$_\mathrm{J}$. Just after the end of their main sequence lifetimes, the irradiance on these planets reached the median incident flux on a typical 1.2 R$_\mathrm{J}$ planet determined by the median incident flux values for confirmed planets listed in the NASA Exoplanet Archive with radii consistent with 1.2 R$_\mathrm{J}$.As the maximum radius of H/He planets determined by structural evolutionary models has been found to be 1.2 R$_\mathrm{J}$, we treat this as the maximum size at which planets could be considered ``uninflated," providing a more conservative incident flux boundary range for inflation than the lower limit established by \citet{demory2011} or the \citet{laughlin2011} planetary effective temperature-radius anomaly models. Now that the host stars have evolved off the main sequence, these planets have reached incident flux values typical for 1.3 R$_\mathrm{J}$ planets. The median incident flux for 1.3 R$_\mathrm{J}$ planets was determined from a sample of confirmed planets taken from the NASA Exoplanet Archive (accessed 9/14/2017).


The average main sequence fluxes of \thestar{b} and \theotherstar{.01} are 170$^{+140}_{-60}$ F$_\oplus$ and 190$^{+150}_{-80}$ F$_\oplus$, respectively. These values are more than 4.5-$\sigma$ from the median fluxes of  well-characterized 1.3 R$_\mathrm{J}$ planets. However, the current incident fluxes of 900$\pm$200 F$_\oplus$ on these planets, shown in green on Figure \ref{evol}, is strongly consistent with the observed incident flux range of 1.3 R$_\mathrm{J}$ planets, suggesting that the radii of these planets is tied closely to their current irradiation. Despite the fact that the planets crossed the empirical threshold for inflation relatively early on in their lifetimes if at all, the planets did not receive sufficient flux to display significant radius anomalies or be inflated to their observed sizes until post-main sequence evolution.

Though the current incident fluxes of the planets in this study lie much closer to the median value for 1.3 R$_\mathrm{J}$ planets, it is important to note that their incident flux is also consistent with the 1.2 R$_\mathrm{J}$ planet population, as the standard deviation in both planet populations is $\gtrsim$500 F$_\oplus$. This is to be expected, as the vast majority of confirmed planet radii are not measured to within 10\% or less, and thus the 1.2 R$_\mathrm{J}$ and 1.3 R$_\mathrm{J}$ planet populations are not distinct.

\subsection{Comparing Re-Inflation and Delayed Cooling Models}

Figure \ref{heating} illustrates Class I models for the radius evolution of \thestar{b} and \theotherstar{.01}, assuming the best-fit values for planet mass, radius, and orbital period. Each of these models assumes a constant planetary heating efficiency, defined to be the fraction of energy a planet receives from its host star that is deposited into the planetary interior, causing adiabatic heating and inflation of the planet. The colors of the various planetary evolution curves correspond to different planetary heating efficiencies ranging from 0.01\% to 0.1\%, assuming a planet with the best-fit planet mass at a constant orbital distance from a star with the best-fit stellar mass calculated here. The incident flux on the planet is then calculated as a function of time using the MESA stellar evolutionary tracks \citep{choi2016}. From this, the planet radius is calculated by convolving the Kelvin-Helmholtz cooling time with planetary heating at a consistent efficiency with respect to the incident stellar flux over the lifetime of the system. The black dotted lines correspond to planetary evolution with no external heat source. Post main sequence evolution is shown with higher time resolution in the insets. Based on the calculated planet radii, we estimate a heating efficiency of 0.03\%$^{+0.04\%}_{-0.02\%}$ for \thestar{b} and 0.03\%$^{+0.03\%}_{-0.01\%}$ for \theotherstar{.01}. Uncertainties on the heating efficiency were calculated by running additional models for each system with both masses and radii lowered/raised by one standard deviation. As planet mass and radius uncertainties are not perfectly correlated, using such a method to calculate planetary heating efficiency should provide conservative errors. 

Based on these two particular planets, the heating efficiency of gas giant planets via post-main sequence evolution of their host stars is strongly consistent between both planets but smaller than theories predict \citep{lopez2016}, and disagrees with the previous estimate of planetary heating efficiency of 0.1\%--0.5\% made by G16. This disagreement stems from the overestimate of the mass of \thestar{b} in the previous study. As the radii of lower density planets are more sensitive to heating and cooling effects than those of higher density, the required heating to inflate a 1.1 M$_\mathrm{J}$ planet to 1.3 R$_\mathrm{J}$ is significantly larger than the heating necessary to inflate a 0.5 M$_\mathrm{J}$ planet to the same size. These new estimates of planet heating efficiency tentatively suggest that if planetary re-inflation occurred in these systems, the process is not as efficient as previous studies suggested \citep{lopez2016}. 

Slowed planetary cooling cannot be entirely ruled out as the cause for large planet radii, as the planets are not larger than they would have been during their pre-main sequence formation. Figure \ref{cooling} illustrates the various delayed cooling tracks that could potentially produce these planets. Different colored curves correspond to cooling models where the Kelvin-Helmholtz cooling time is increased by a constant factor. \thestar{b} and \theotherstar{.01} are shown in bold, whereas planets with masses of 0.4--0.6 M$_\mathrm{J}$, incident fluxes of 100--300 F$_\oplus$, and host stars smaller than 2R$_\odot$ (to ensure that they have not begun RGB evolution) are shown in gray (specifically these planets are K2-30b, Kepler-422b, OGLE-TR-111b, WASP-11b, WASP-34b, and WASP-42b). It can be seen that the main sequence planets have systematically smaller radii, and thus suggest delayed cooling rates that are significantly different from those which would be inferred from the planets in this study. The required cooling delay factor for the post-main sequence planets studied here is 20--250, significantly more than the factor of $\sim$1--10 for main sequence cases. Delayed cooling models predict a decrease in planet radius with age, which strongly disagrees with the data shown here. Re-inflation models predict the opposite. Thus, we conclude that Class I re-inflation mechanisms are more statistically relevant than Class II mechanisms in the evolution of \thestar{b} and \theotherstar{.01}, and thus stellar irradiation is likely to be the direct cause of warm and hot Jupiter inflation.


Furthermore, the assumption of a 10M$_\oplus$ core is low compared to the inferred core masses of cooler non-inflated giants. Using the planet-core mass relationship of \citet{thorngren2015}, we predict core masses of $\approx$37 M$_\oplus$ for both \thestar{b} and \theotherstar{.01}. These higher core masses would significantly increase the required heating efficiencies to 0.10\%$^{+0.09\%}_{-0.05\%}$ for \thestar{b} and 0.14\%$^{+0.07\%}_{-0.04\%}$ for \theotherstar{.01}, or delayed cooling factors of 300--3000$\times$ for these planets. Though these values suggest better agreement with previous results (e.g., G16), we report the conservative outcomes assuming 10 M$\oplus$ cores to place a lower limit on the efficiency of planetary heating.


\subsection{Eccentricity Effects}

\citet{jones2017} independently report a non-zero eccentricity for \theotherstar{.01} based on the HIRES data presented here and additional RV measurements obtained with other instruments. Since transit parameters are often degenerate, an inaccurate eccentricity could result in an inaccurate planet radius \citep[e.g.][]{eastman2013} and thus potentially affect our conclusions regarding planet re-inflation.

A non-circular orbit would be surprising given the expected tidal circularization timescale for such planets. Our estimated planet parameters suggest a timescale of $\tau_e \sim$ 6 Gyr using the relation of \citet{gu2002} and assuming a tidal quality factor $Q_p \approx$ 10$^6$, comparable to Jupiter \citep{ogilvie2004, wu2005}. This suggests that the orbit of this planet should have been circularized before post-main sequence evolution, as long as no other companion could have dynamically excited the system. However, these timescale estimates are very sensitive to planet density and tidal quality factor, and adjusting these parameters within errors can result in estimates of $\tau_e <$ 1 Gyr as well as $\tau_e >$ 10 Gyr. Thus, we cannot rule out a non-zero eccentricity for this system based on tidal circularization timescale arguments alone. 

We also used the relations of \citet{bodenheimer2001} to determine the tidal circularization energy and thus tidal radius inflation that would expected for this system. We find that the tidal inflation should be negligible for this system even for a potentially high eccentricity. Thus, if this planet were to be on an eccentric orbit, we should still be able to distinguish between tidal and irradiative planet inflation.


We attempted to model the eccentricity of this system and obtained results which were consistent with our circular model. However, these tests resulted in non-convergent posterior chains, and thus we cannot rule out a non-negligible eccentricity for this system. Additional RV measurements should help to constrain the eccentricity of this system, and clarify if and how eccentricity affects the planet radius presented here.

\subsection{Selection Effects and the Similarity of Planet Parameters}

\thestar{} and \theotherstar{} are remarkably similar:  the stellar radii and masses and planet radii, masses, and orbital periods agree within 10\%. This begs the question: is it only coincidence that these systems are so similar, is it the product of convergent planetary evolution, or is it the result of survey bias or selection effect?  Here, we investigate the last possibility.

Two effects modulate the intrinsic distribution of planets as a function of mass $M$, radius $R$, and orbital period $P$ to produce the observed occurrence in a survey of evolved stars: the detection of the planet by transit, and the lifetime of planets against orbital decay due to tides raised on large, low-density host stars.  A deficit of giant planets close to evolved stars \citep{Kunitomo2011} as well as the peculiar characteristics of some RGB stars (rapid rotation, magnetic fields, and lithium abundance) have been explained as the result of orbital decay and ingestion of giant planets \citep{Carlberg2009,Privitera2016a,Privitera2016b,Aguilera2016a,Aguilera2016b}.

The volume $V$ over which planets of radius $R_p$ and orbital period $P$ can be detected transiting a star of mass and radius $M_*$ and $R_*$ is (see Appendix): 
\begin{equation}
\label{eqn:volume}
V \sim R_p^{\frac{3}{(1-\alpha)}}P^{-1}R_*^{-\frac{3(3\alpha-1)}{2(1-\alpha)}}M_*^{-\frac{1}{2}},
\end{equation}.
where $\alpha$ is the power-law index relating RMS photometric error to number of observations ($\alpha = 1/2$ for uncorrelated white noise).  The lifetime of a planet against orbital decay due to tides raised on the star, in the limit that the decay time is short compared to the RGB lifetime, is 
\begin{equation}
\label{eqn:lifetime}
\tau_{\rm tide} \approx 4.1\left(\frac{M_P}{M_J}\right)^{-1}P_{\rm days}^{\frac{13}{3}}\frac{Q'_*}{2 \times 10^5}\left(\frac{M_*}{M_{\odot}}\right)^{\frac{5}{3}}\left(\frac{R_*}{R_{\odot}}\right)^{-5}\,{\rm Myr},
\end{equation}.
where $Q'_*$ is a modified tidal quality factor (see Appendix).

The bias effect $B$ is the product $V \cdot \tau_{\rm tide}$ which then scales as:
\begin{equation}
\label{eqn.bias}
B \propto R_p^{\frac{3}{(1-\alpha)}}M_P^{-1}P^{\frac{10}{3}}M_*^{\frac{7}{6}}R_*^{-\frac{7-\alpha}{2(1-\alpha)}}.
\end{equation}
This formulation ignores the possibility of Roche-lobe overflow and mass exchange between the planet and the star \citep[e.g.,][and references therein]{Jackson2017}.  Roche-lobe overflow of the planet will occur only when $a \lesssim 2.0 R_* (\rho_*/\rho_p)^{1/3}$ \citep{Rappaport2013} and since $\rho_p$ is at least an order of magnitude larger than $\rho_*$ on the RGB, overflow never occurs before the planet is engulfed.  In fact, the planet may accrete mass from the star before engulfment but this only hastens its demise.  

\begin{figure}
\centering
\includegraphics[width=0.5\textwidth]{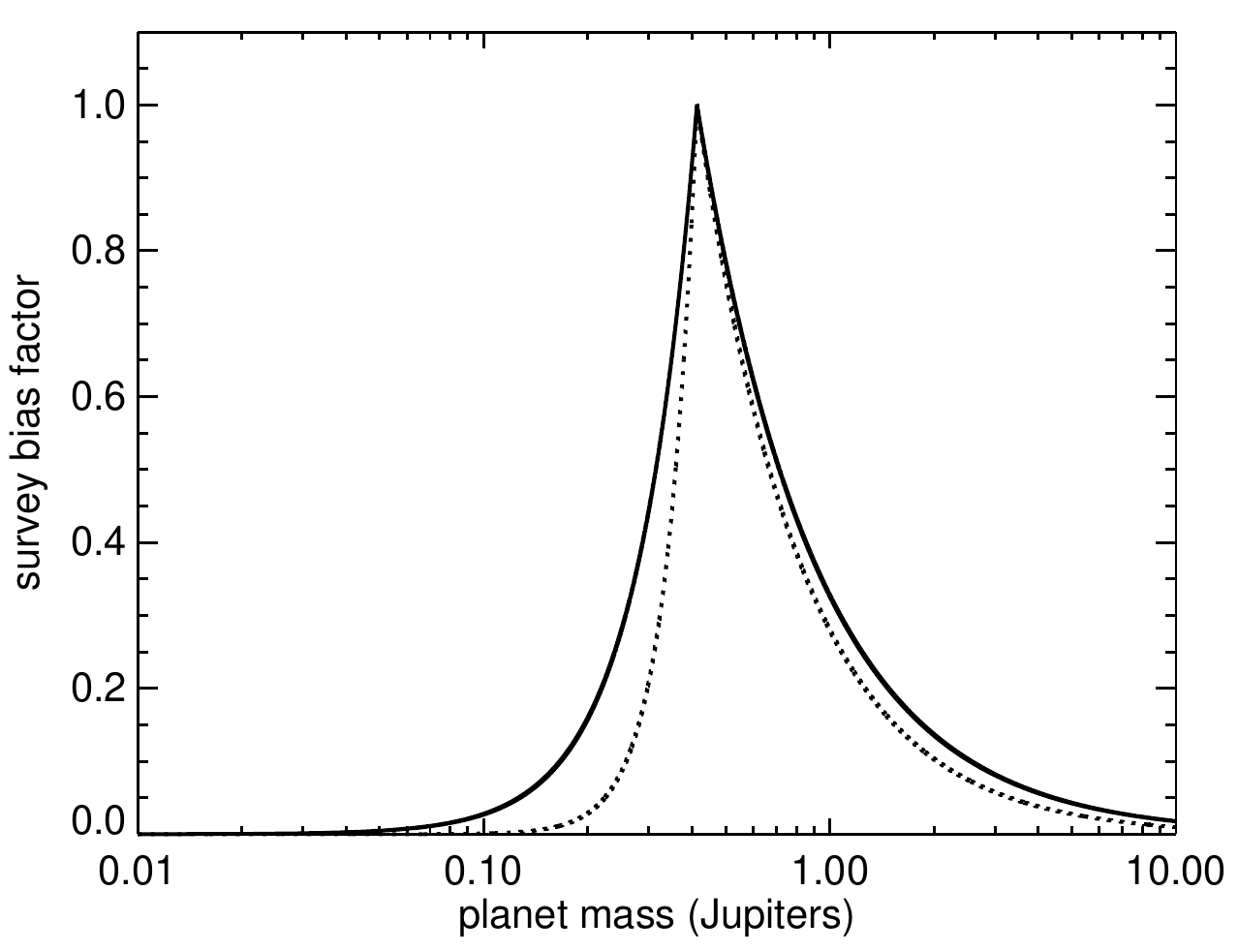}
\caption{Survey bias factor $B$ as a function of planet mass for planets around evolved stars, calculated using Eqn. \ref{eqn.bias} and the \citet{chen2017} planet mass-radius relation, and assuming the orbital decay time is much shorter than the stellar evolution time.  The solid lines is for pure ``white" (Poisson) noise ($\alpha = 0.5$) while the dashed line is for the case of ``red" (correlated) noise ($\alpha = 0.7$).  Detection of planets of 0.4$M_J$ mass is strongly favored: smaller planets are more difficult to detect while more massive planets do not survive long enough.}
\label{fig.bias}
\end{figure}

Our survey is biased towards planets with large radii (easier to detect) but against planets with large masses (shorter lifetime).  Contours of constant bias in a mass-radius diagram describe the relation $R_P \propto M_P^{(1-\alpha)/3}$.  If the power-law index of the planetary mass-radius relation is \emph{steeper} than the critical value $(1-\alpha)/3$ then larger planets are favored; if it is shallower than smaller planets are favored.     A maximum in $B$ occurs where the index breaks, i.e. at a ``knee" in the mass-radius relation.  For $\alpha = 1/2$ the critical value of the power-law index is $1/6$, i.e. well below the values inferred for rocky planets or ``ice giants" like Neptune.   \citet{chen2017} inferred a break at $0.41 \pm 0.06M_J$ where the index falls from 0.59 to -0.04, reflecting the onset of support by electron degeneracy in gas giant planets.  \citet{Bashi2017} found a similar transition of 0.55 to 0.01 at $0.39 \pm 0.02M_J$. Since the power-law index of $B$ is bounded by 0 and 1/3, the location of $B$ is independent of $\alpha$, but the magnitude of the bias does increase with $\alpha$.  This is illustrated in Fig. \ref{fig.bias}, where $B$ (normalized by the maximum value) is calculated for planets following the \citet{chen2017} mass-radius relation and with $\alpha = 1/2$ (pure Poisson noise) and $\alpha = 0.7$ (finite correlated noise).  

For periods less than a critical value $P_*$ (see Appendix), where
\begin{equation}
\label{eqn.pone}
P_* = 0.63 \left(M_P \tau_{\rm RGB} M_*^{-1}\rho_i^{-5/3}\right)^{3/13}\,{\rm days},
\end{equation}
where $M_P$ is in Jupiter masses, $\tau_{\rm RGB}$ is in Myr, and $M_*$ and $\rho_i$ are in solar units, the decay time is shorter than the RGB lifetime and Eqn. \ref{eqn.pone} holds.  Using the stellar evolution models of \citet{Pols1998} for a solar-like metallicity, we find $P_* \approx 5-6$~days, roughly independent of $M_*$ over the range 0.9--1.6\msun, and only weakly dependent on $M_P$.  For planets with $P>P_*$, including \thestar{b} and \theotherstar{.01}, planet lifetime is governed by the RGB evolution time rather than orbital decay time, and detection bias dominates.\\ 

The survey bias for $P$ can be seen in Eqn. \ref{eqn.bias} where $B$ increases rapidly with $P$ to $P_*$, at which point $\tau_{\rm tide}$ becomes comparable to $\tau_{\rm RGB}$ and Eqn. \ref{eqn.bias} no longer applies.  Beyond that point, survey bias is governed by detection bias, which \emph{decreases} with $P$ (Eqn. \ref{eqn:volume}).  Thus $B$ has a maximum at $P = 5-6$~days, weakly dependent on planet mass and $Q_*$.  This potentially can explain Kepler-91b (6.25~days), but perhaps not \thestar{b} or \theotherstar{.01}.  

Since $P_*$ is weakly $M_P$-dependent, survey bias at $P=P_*$ is also dependent on both $R_P$ and $M_P$.  Substituting Eqn. \ref{eqn.pone} into Eqn. \ref{eqn.bias} yields $B \propto R_P^{3/(1-\alpha)}M_P^{-3/13}$.  Interestingly, this mass dependence, combined with the slightly negative mass-radius power-law index for giant planets due to electron degeneracy pressure, is enough to produce a peak in $B$, again at the $0.4M_J$ transition.   Explanation of the similarities of the \thestar{b} and \theotherstar{.01} systems by survey bias, however, might require an anomalously low value of $Q'_*$, inconsistent with constraints from binary stars and analyses of other planetary systems \citep[see discussion in ][]{patra2017}, as well as the theoretical expectation that dissipation on the RGB is weaker because of the small core mass and radius \citep[e.g.,][]{Gallet2017}.  

Alternatively, we note that our selection criterion criterion of detectable stellar oscillations imposes a lower limit on $R_*$ of about 3\rsun.  This means that that effective initial stellar density in our sample $\rho_i$ is several times smaller, which increase $P_*$ by a factor of $\sim1.5$, making it consistent with the orbits of \thestar{b} and \theotherstar{.01}.  In future work we will perform a more rigorous treatment of bias using the actual stars in our survey and their properties using asteroseismology, spectroscopy, and forthcoming \emph{Gaia} parallaxes.

\section{Conclusions}


We report the discovery of a transiting planet with R = \otherplanrad{} and M = \otherplanmass{} around the low luminosity giant star \theotherstar, and revise our earlier mass estimate of \thestar{b}. We use a simple harmonic oscillator Gaussian process model to estimate the correlated noise in the lightcurve to quantify and remove potential correlations between planetary and stellar properties, and measure asteroseismic quantities of the star using only time domain information. We also performed  spectroscopic, traditional asteroseismic, and imaging studies of the host stars \thestar{} and \theotherstar{} to precisely determine stellar parameters and evolutionary history and rule out false positive scenarios. We find that both systems have effectively null false positive probabilities. We also find that the masses, radii, and orbital periods of these systems are similar to within 10\%, possibly due to a selection bias toward larger yet less massive planets.

We determine that \thestar{b} and \theotherstar{.01} require approximately 0.03$\%$ of the current incident stellar flux to be deposited into the planets' deep convective interior to explain their radii. This suggests planet inflation is a direct response to stellar irradiation rather than an effect of delayed planet cooling after formation, especially for inflated planets seen in evolved systems. However, stellar irradiation may not be as efficient a mechanism for planet inflation as indicated by \citet{grunblatt2016}, due to the previously overestimated mass of \thestar{b} driven by the limited phase coverage of the original Keck/HIRES radial velocity measurements.


Further studies of planets around evolved stars are essential to confirm the planet re-inflation hypothesis. Planets may be inflated by methods that are more strongly dependent on other factors such as atmospheric metallicity than incident flux. An inflated planet on a 20~day orbit around a giant star would have been definitively outside the inflated planet regime when its host star was on the main sequence, and thus finding such a planet could more definitively test the re-inflation hypothesis. Similarly, a similar planet at a similar orbital period around a more evolved star will be inflated to a higher degree (assuming a constant heating efficiency for all planets). Thus, discovering such a planet would provide more conclusive evidence regarding these phenomena. Heating efficiency may also vary between planets, dependent on composition and other environmental factors. Continued research of planets orbiting subgiant stars and planet candidates around larger, more evolved stars should provide a more conclusive view of planet re-inflation. 

The NASA TESS Mission \citep{sullivan2015} will observe over 90\% of the sky with similar cadence and precision as the K2 Mission for 30 days or more. This data will be sufficient to identify additional planets in $\sim$10 day orbital periods around over an order of magnitude more evolved stars, including oscillating red giants \citep{campante2016}. This dataset should be sufficient to constrain the heating efficiency of gas-giant planets to the precision necessary to effectively distinguish between delayed cooling and direct re-inflationary scenarios. It will also greatly enhance our ability to estimate planet occurrence around LLRGB stars and perhaps help determine the longevity of our own planetary system.

\acknowledgements{The authors would like to thank Jonathan Fortney, Ruth Angus, Ashley Chontos, Travis Berger, Allan Simeon, Jr., Jordan Vaughan, and Stephanie Yoshida for helpful discussions. This research was supported by NASA Origins of Solar Systems grant NNX11AC33G to E.G. and by the NASA K2 Guest Observer Award NNX16AH45G to D.H..  D.H. acknowledges support by the Australian Research Council's Discovery Projects funding scheme (project number DE140101364) and support by the National Aeronautics and Space Administration under Grant NNX14AB92G issued through the Kepler Participating Scientist Program. W.J.C. and T.S.H.N. acknowledge support from the UK Science and Technology Facilities Council (STFC). A.V. is supported by the NSF Graduate Research Fellowship, grant No. DGE 1144152. This research has made use of the Exoplanet Orbit Database and the Exoplanet Data Explorer at Exoplanets.org. This work benefited from the Exoplanet Summer Program in the Other Worlds Laboratory (OWL) at the University of California, Santa Cruz, a program funded by the Heising-Simons Foundation. This work was based on observations at the W. M. Keck Observatory granted by the University of Hawaii, the University of California, and the California Institute of Technology. We thank the observers who contributed to the measurements reported here and acknowledge the efforts of the Keck Observatory staff. We extend special thanks to those of Hawaiian ancestry on whose sacred mountain of Maunakea we are privileged to be guests. Some/all of the data presented in this paper were obtained from the Mikulski Archive for Space Telescopes (MAST). STScI is operated by the Association of Universities for Research in Astronomy, Inc., under NASA contract NAS5-26555. Support for MAST for non-HST data is provided by the NASA Office of Space Science via grant NNX09AF08G and by other grants and contracts. This research has made use of the NASA/IPAC Infrared Science Archive, which is operated by the Jet Propulsion Laboratory, California Institute of Technology, under contract with the National Aeronautics and Space Administration. This research made use of the SIMBAD and VIZIER Astronomical Databases, operated at CDS, Strasbourg, France (http://cdsweb.u-strasbg.fr/), and of NASAs Astrophysics Data System, of the Jean-Marie Mariotti Center Search service (http://www.jmmc.fr/searchcal), co-developed by FIZEAU and LAOG/IPAG. E.D.L. received funding from the European Union Seventh Framework Programme (FP7/2007- 2013) under grant agreement number 313014 (ETAEARTH). Any opinion, findings, and conclusions or recommendations expressed in this material are those of the authors and do not necessarily reflect the views of the National Science Foundation.}

\bibliography{reinfpaper}

\appendix{}

\subsection{Survey Bias for Star and Planet Properties}


Following \citet{gaudi2005}, we estimated the distance $d$ to which systems can be detected, but we modify the calculation to account for coherent (``red") noise from stellar granulation and noise due to drift of the spacecraft and stellar image on the \ktwo{} CCDs, whereby the RMS noise increases faster than the the square root of the number of measurements $n$, or the signal-to-noise decreases more slowly than $n^{-1/2}$.  We parameterize this by the index $\alpha$, where the RMS noise scales as $n^{\alpha}$.  In a magnitude-limited survey of stars of a monotonic color (i.e. bolometric correction) and fixed solid angle, the volume $V$ that can observed to a distance $d$ and hence the number of systems in a survey goes as $d^3$.  This scales as\footnote{This assumes that $d$ does not extend outside the galactic disk over a significant portion of the survey.}:
\begin{equation}
V \propto R_p^{\frac{3}{(1-\alpha)}}P^{-1}R_*^{-\frac{3\alpha}{1-\alpha}}\rho_*^{-\frac{1}{2}}.
\end{equation}
For the case of $\alpha = 1/2$ (white noise) we recover the original scaling of \citet{gaudi2005}:
\begin{equation}
V \propto R_p^{6}R_*^{-\frac{3}{1-\alpha}}P^{-1}\rho_*^{-\frac{1}{2}}.
\end{equation}  
Since stars on the RGB differ far more in radius than they do in mass, we re-express $\rho_*$ in Eqn. \ref{eqn:volume} terms of $M_*$ and $R_*$:
\begin{equation}
V \sim R_p^{\frac{3}{(1-\alpha)}}P^{-1}R_*^{-\frac{3(3\alpha-1)}{2(1-\alpha)}}M_*^{-\frac{1}{2}}
\end{equation}

We also consider the lifetime of a planet against orbital decay due to the tides it raises on the slowly-rotating star.  This is expressed as \citep[e.g.,][]{patra2017}:
\begin{equation}
\label{eqn.orbitevo}
\frac{dP}{dt} = - \frac{27\pi}{2Q'_*}\frac{M_P}{M_*}\left(\frac{3\pi}{G\rho_*}\right)^{\frac{5}{3}}P^{-\frac{10}{3}},
\end{equation}
where $Q'_*$ is a modified tidal dissipation factor that includes the Love number, $M_*$ and $\rho_*$ the stellar mass and mean density, and $G$ is the gravitational constant.  

If a planet's orbit decays on a time scale that is short compared to any evolution of the host star on the RGB (i.e. $R_*$ is constant) and mass loss is negligible (i.e. $M_*$ is constant) then integrating Eqn. \ref{eqn.orbitevo} yields the decay lifetime $\tau_{\rm tide}$:
\begin{equation}
\tau_{\rm tide} \approx 4.1\left(\frac{M_P}{M_J}\right)^{-1}P_{\rm dy}^{\frac{13}{3}}\frac{Q'_*}{2 \times 10^5}\left(\frac{M_*}{M_{\odot}}\right)^{\frac{5}{3}}\left(\frac{R_*}{R_{\odot}}\right)^{-5}\,{\rm Myr},
\end{equation}
where stellar values are those at the base of the RGB.  

\begin{figure}
\centering
\includegraphics[width=0.5\textwidth]{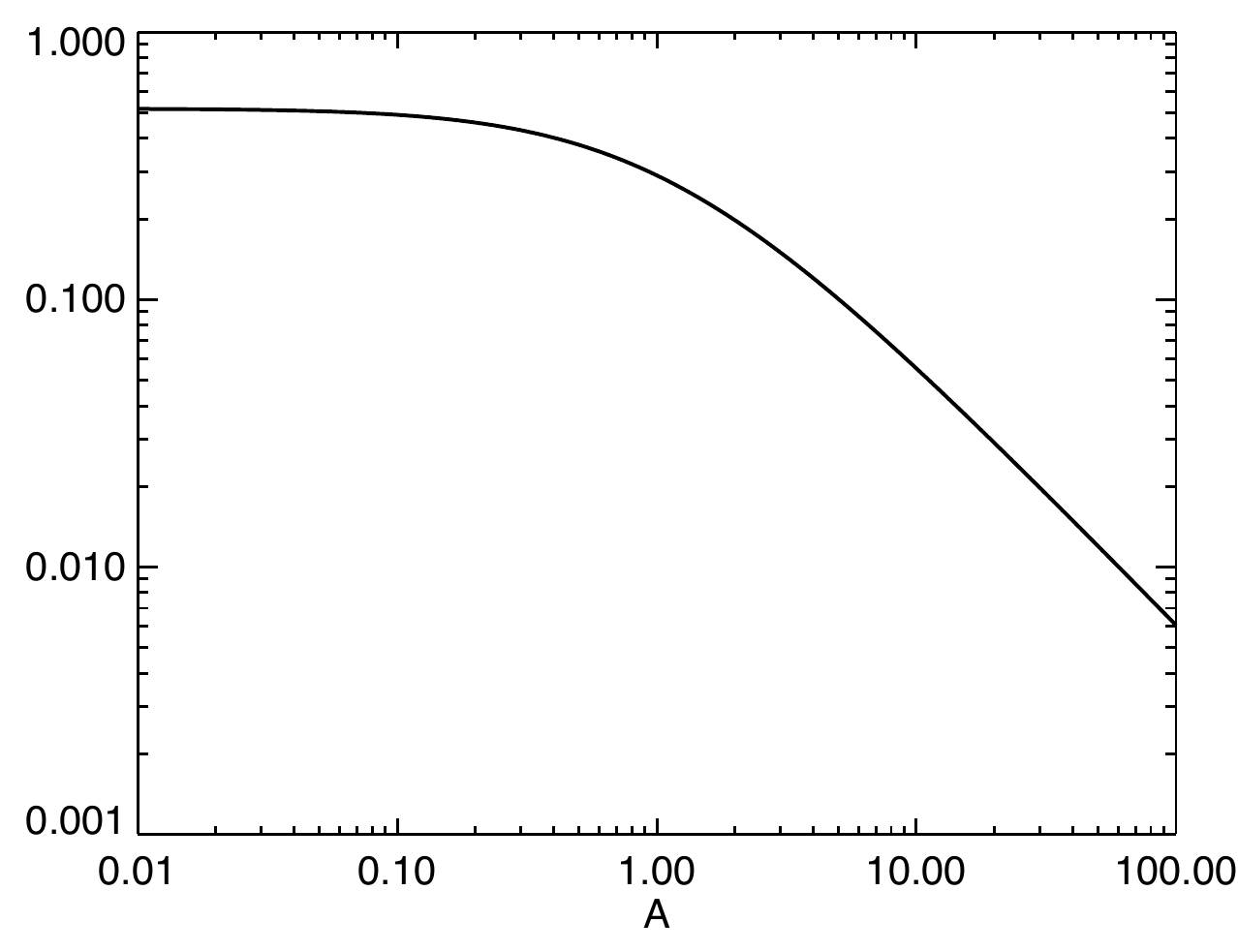}
\caption{Survey bias factor $B$ as a function of $A$ (Eqn. \ref{eqn.a}), which contains the dependencies on $M_P$, $P$, and $R_*$, and accounts for simultaneous orbital decay and evolution of the host star along the RGB.  In the regime where $A \gg 1$ (orbital decay faster than stellar evolution), $B \propto 1/A$ and Eqn. \ref{eqn.bias} is recovered.  If $A \ll 1$, $B$ is independent of $A$ and dependent only on $R_P$.}
\label{fig.biastwo}
\end{figure}

For sufficiently low $M_P$ or large $P$ the orbital decay time becomes comparable to the timescale of evolution of the host star on the RGB.  $R_*$ increases, decreasing the volume over which the planet could be detected (Eqn. \ref{eqn:volume}), and shortens the lifetime (Eqn. \ref{eqn:lifetime}).  Rather than $V \tau_{\rm tide}$, we must evaluate
\begin{equation}
\label{eqn:btwo}
B \propto \int_0^{\tau_{\rm tide}} dt\,V(t).
\end{equation}
To model the density evolution on the RGB during H-shell burning we adopt a helium core-mass evolution equation:
\begin{equation}
\frac{dM_c}{dt} = -\frac{L}{X\xi},
\end{equation}
where $L$ is the luminosity, $X$ is the mixing ratio of H fuel ($\approx 0.7$) and $\xi$ the energy release for H-burning.  We use the core mass-luminosity relation of \citet{Refsdal1970}:
\begin{equation}
\frac{L}{L_{\odot}}\approx 200 \left(\frac{M_c}{M_0}\right)^{\beta}
\end{equation}
where $M_0 = 0.3$\msun{} is a reference core mass and $\beta = 7.6$.  Assuming a constant \teff{} so that $L_* \propto R_*^2$ and neglecting mass loss on the RGB, the density evolves as;
\begin{equation}
R_* = R_i \left[1-\frac{L_0\left(\beta-1\right)}{M_0 X\xi}\left(\frac{M_i}{M_0}\right)^{\beta-1}\right]^\frac{-\beta}{2(\beta-1)},
\end{equation}
where $\rho_i$ and $M_i$ are the initial stellar density and core mass on the RGB.  This can be re-written in terms of the duration of the RGB phase $\tau_{\rm RGB}$ and the final core mass $M_f$ at the tip of the RGB when the helium flash occurs:
\begin{equation}
\label{eqn:radius}
R_*(t) = R_i \left[1 + \frac{t}{\tau_{\rm RGB}}\left[1 - \left(\frac{M_i}{M_f}\right)^{\beta-1}\right]\right]^{\frac{-\beta}{2(\beta-1)}}
\end{equation}
By the time the helium flash occurs, the radius of the star has evolved considerably, i.e. $R_f/R_i = \left(M_f/M_i\right)^{\beta/2}$.  For a solar-mass star, $M_f/M_i \approx 4$ \citep{Pols1998} and stars at the RGB tip will have enlarged by over two orders of magnitude relative to the end of the main sequence, while $\tau_{\rm tide}$ will have fallen by a factor of $10^{11}$ (Eqn. \ref{eqn:lifetime}).  We assume that the no planet of interest survives that long, i.e. $\tau_{\rm tide}$ never approaches $\tau_{\rm RGB}$.  Moreover, even giant planets will not be detected by transit because $R_P/R_*$ will be too small, and we neglect the mass term in Eqn. \ref{eqn:radius}:
\begin{equation}
R_*(t) \approx R_i \left(1 - \frac{t}{\tau_{\rm RGB}}\right)^{\frac{-\beta}{2(\beta-1)}}
\end{equation}
To obtain a scaling relation for $\tau_{\rm tide}$ we substitute Eqn. \ref{eqn:radius} into Eqn. \ref{eqn.orbitevo} to and integrate to obtain $P(t)$, then evaluate the time-dependent factors in Eqn. \ref{eqn:btwo}.  Substituting $x = 1-t/\tau_{\rm RGB}$, $B$ scales as
\begin{equation}
\begin{aligned}
B\; \propto \; & R_p^{\frac{3}{1-\alpha}}M_*^{-\frac{1}{2}}\tau_{\rm RGB}\\
& \times \int_{x_{\rm min}}^1 dx\;\left[1 - A\left(x^{-\frac{3\beta + 2}{2(\beta-1)}}-1\right)\right]^{-\frac{3}{13}} x^{\frac{3\beta (3\alpha-1)}{4(1-\alpha)(\beta-1)}},
\end{aligned}
\end{equation}
where 
\begin{equation}
\label{eqn.a}
A = \frac{117\pi}{Q_*}\frac{\beta-1}{3\beta+2}\frac{M_P}{M_*}\left(\frac{3\pi}{G\rho_i}\right)^{5/3}\frac{\tau_{\rm RGB}}{P_0^{13/3}},
\end{equation}
and
\begin{equation}
x_{\rm min} = \left(1 + A^{-1}\right)^{-\frac{2(\beta-1)}{3\beta+2}}.
\end{equation}
Figure \ref{fig.biastwo} plots $B$ as a function of $A$ for $\beta = 7.6$ and $\alpha = 1/2$.  It shows that if $A \gg 1$ (rapid tidal evolution) then $B \propto A^{-1}$ and hence $B \propto R_P^{3/(1-\alpha)}M_p^{-1}$, as in Eqn. \ref{eqn.bias} and thus detection of transition objects at the electron degeneracy threshold is favored.  However, if $A \ll 1$ then $B$ is independent of $A$ and hence $M_P$ and $P$ (but not $R_P$).  Detection of gas giants, particularly inflated planets with the largest radii, is then favored.  For the same values of $\alpha$ and $\beta$ and $Q_* = 2 \times 10^5$, the condition for $A=1$ becomes a critical value for period
\begin{equation}
P_* = 0.63 \left(M_P \tau_{\rm RGB} M_*^{-1}\rho_i^{-5/3}\right)^{3/13}\,{\rm days},
\end{equation}
where $M_P$ in Jupiter masses, $\tau_{\rm RGB}$ is in Myr, and $M_*$ and $\rho_i$ are in solar units.

\end{document}